\documentclass{journalA4}

\usepackage{graphicx,amssymb,amsmath,fancyhdr,paralist,xspace,euscript,picins}
\usepackage{wrapfig, subfig,algorithmic}
\usepackage[usenames]{color}
\newcommand{\twist}{twist}

\newcommand{\LineY}[2]{\overline{~#1 #2~}}
\newcommand{\gadget}[1]{\ensuremath{g_{#1}}}

\newcommand{\hLine}[1]{H_{#1}}

\newcommand{\vtx}{\mathsf{v}}

\newcommand{\splitVtx}[2]{\vtx_{#1}^{#2}}
\newcommand{\splitCoordV}[2]{v_{#1}^{#2}}
\newcommand{\splitEdgeYes}[2]{\edge_{#1}^{#2}}
\newcommand{\splitEdgeNo}[2]{\overline{\edge}_{#1}^{#2}}
\newcommand{\splitCoordA}[2]{a_{#1}^{#2}}
\newcommand{\splitPntA}[2]{\pntA_{#1}^{#2}}
\newcommand{\splitCoordB}[2]{b_{#1}^{#2}}
\newcommand{\splitPntB}[2]{\pntB_{#1}^{#2}}
\newcommand{\splitCoordC}[2]{c_{#1}^{#2}}
\newcommand{\splitPntC}[2]{\pntC_{#1}^{#2}}
\newcommand{\splitCoordD}[2]{d_{#1}^{#2}}
\newcommand{\splitPntD}[2]{\pntD_{#1}^{#2}}
\newcommand{\splitCoordP}[2]{p_{#1}^{#2}}
\newcommand{\splitPntP}[2]{\pntP_{#1}^{#2}}

\newcommand{\splitPntBi}[1]{\pntB_{#1}}

\newcommand{\scCurve}[1]{#1_{\Diamond}}
\newcommand{\traversal}{feasible shortcut curve}

\newcommand{\ZoneVar}{\zeta}
\newcommand{\MonotoneConst}{4}
\newcommand{\DistanceProjection}{Distance Projection}
\newcommand{\baseY}{base curve}
\newcommand{\targetX}{target curve}

\newcommand{\edge}{\mathsf{e}}
\newcommand{\SubCrv}[3]{{#1}\pbrcZ{#2, #3}}

\newcommand{\ScutCrv}[3]{\overline{#1}\!\pbrcZ{#2, #3}}

\newcommand{\pbrcZ}[1]{\left[ \MakeSBig {#1} \right]}
\providecommand{\eps}{{\varepsilon}}
\renewcommand{\Re}{{\rm I\!\hspace{-0.025em} R}}
\providecommand{\MakeSBig}{\rule[0.0cm]{0.0cm}{0.35cm}} % really small
\providecommand{\pth}[2][\!]{#1\left({#2}\right)}
\newcommand{\subcX}[1]{\SubCurvifyY{\cX}_{#1}}
\newcommand{\cX}{\SimplifyX{\cXBase}}
\newcommand{\cY}{\SimplifyX{\cYBase}}
\newcommand{\cXBase}{T}
\newcommand{\cYBase}{B}
\newcommand{\SimplifyX}[1]{#1}
\newcommand{\x}{x}
\newcommand{\y}{y}
\newcommand{\Frechet}{Fr\'{e}c{h}e{}t\xspace}
\newcommand{\pnt}{\mathsf{p}}
\newcommand{\distFr}[2]{\mathsf{d}_{\EuScript{F}}\pth{#1, #2}}
\newcommand{\distX}[2]{\distCmd{#1 - #2}}
\newcommand{\distCmd}[1]{\left\| {#1} \right\|}
\newcommand{\distSFr}[2]{\mathsf{d}_{\EuScript{S}}\pth{#1,#2}}
\newcommand{\pntA}{\mathsf{a}}
\newcommand{\pntB}{\mathsf{b}}
\newcommand{\pntC}{\mathsf{c}}
\newcommand{\pntD}{\mathsf{d}}
\newcommand{\pntP}{\mathsf{p}}
\newcommand{\pntQ}{\mathsf{q}}

\newcommand{\vertex}{\mathsf{v}}
\newcommand{\Array}{\EuScript{A}}

\newcommand{\subsegY}[1]{\overline{\cY}_{#1}}

\newcommand{\SubCurvifyY}[1]{{#1}}
\renewcommand{\th}{t{}h\xspace}
\newcommand{\frVal}[1]{\mathsf{d}\pth{#1}}
\newcommand{\PntSet}{{\mathsf{P}}}

\newcommand{\ACoord}[1]{\x_{#1}}
\newcommand{\BCoord}[1]{\y_{#1}}
\newcommand{\xPnt}{\ACoord{\pnt}}
\newcommand{\yPnt}{\BCoord{\pnt}}

\newcommand{\xPntQ}{\ACoord{\pntQ}}
\newcommand{\yPntQ}{\BCoord{\pntQ}}

\newcommand{\ReachFDleqI}[1]{\ensuremath{\EuScript{R}_{\leq #1}}}
\newcommand{\RCellXY}[3]{\ReachFDleqI{#1}^{(#2,#3)}}

\newcommand{\FullFDleqC}{\EuScript{D}}
\newcommand{\FullFDleq}[1]{\FullFDleqC_{\leq #1}}
\newcommand{\FCellXY}[3]{\FullFDleq{#1}^{(#2,#3)}}

\providecommand{\pth}[2][\!]{#1\left({#2}\right)}
\providecommand{\brc}[1]{\left\{ {#1} \right\}}
\providecommand{\sep}[1]{\,\left|\, {#1} \MakeBig\right.}
\providecommand{\MakeBig}{\rule[-.2cm]{0cm}{0.4cm}}

\newcommand{\Cell}{C}
\newcommand{\CellXY}[2]{\Cell_{#1,#2}}

\providecommand{\MakeBig}{\rule[-.2cm]{0cm}{0.4cm}}
\providecommand{\MakeSBig}{\rule[0.0cm]{0.0cm}{0.35cm}} % really small

\newcommand{\RVCellXY}[2]{R^v_{#1,#2}}
\newcommand{\RHCellXY}[2]{R^h_{#1,#2}}

\newcommand{\tunnelLtr}{\mathsf{\tau}}
\newcommand{\xtunnel}[2]{\tunnelLtr\pth{ #1,  #2}}

\newcommand{\scPrice}[2]{{\mathrm{p{r}c}}_{\tunnelLtr}\pth{ #1, #2}}

\newcommand{\Interval}{\mathcal{I}}

\definecolor{blue25}{rgb}{0,0,0.55}

\newcommand{\seclab}[1]{\label{sec:#1}}
\newcommand{\secref}[1]{Section~\ref{sec:#1}}
\newcommand{\lemlab}[1]{\label{lem:#1}}
\newcommand{\lemref}[1]{Lemma~\ref{lem:#1}}
\newcommand{\figlab}[1]{\label{fig:#1}}
\newcommand{\figref}[1]{Figure~\ref{fig:#1}}
\newcommand{\defref}[1]{Definition~\ref{def:#1}}

\newcommand{\corlab}[1]{\label{cor:#1}}
\newcommand{\corref}[1]{Corollary~\ref{cor:#1}}
\newcommand{\obslab}[1]{\label{obs:#1}}
\newcommand{\obsref}[1]{Observation~\ref{obs:#1}}

\newtheorem{corollary}[theorem]{Corollary}

\newcommand{\vrestricted}{vertex-restricted}

\newcommand{\unrestricted}{continuous}
\newcommand{\asymmetric}{directed}
\newcommand{\symmetric}{undirected}

\newcommand{\distoSFr}[2]{d_{\EuScript{S}}\pth{#1,#2}}

\newcommand{\pbrc}[2][\!\!]{#1\left[ {#2} \MakeBig \right]}

\pdfoutput=1

\title{Computing the \Frechet distance with shortcuts is NP-hard}

\author{Maike Buchin\thanks{Faculty of Mathematics,
Ruhr-University Bochum, Germany, {\tt Maike.Buchin@rub.de.}}
\and Anne Driemel\thanks{ Department of Computer Science, TU Dortmund, Germany, {\tt
anne.driemel@udo.edu}. Work on this paper was partially supported by the Netherlands Organisation for
Scientific Research (NWO) under project no. 612.065.823.}
\and Bettina Speckmann\thanks{Department of Mathematics and Computer Science,
TU Eindhoven, the Netherlands, {\tt speckman@win.tue.nl}. Supported by
the Netherlands Organisation for Scientific Research (NWO) under project no. 639.022.707.}}

\date{}

\begin{document}
\maketitle

\begin{abstract}\noindent
%short version:
We study the \emph{shortcut \Frechet distance}, a natural variant of the
\Frechet distance, that allows us to take shortcuts from and to any point along
one of the curves.  The classic \Frechet distance is a bottle-neck distance
measure and hence quite sensitive to outliers. The shortcut \Frechet distance
allows us to cut across outliers and hence produces more
meaningful results when dealing with real world data.  Driemel and Har-Peled
recently described approximation algorithms for the restricted case where
shortcuts have to start and end at input vertices.  We show that, in the
general case, the problem of computing the shortcut \Frechet distance is
NP-hard.  This is the first hardness result for a variant of the \Frechet
distance between two polygonal curves in the plane.  We also present two
algorithms for the decision problem: a $3$-approximation algorithm for the
general case and an exact algorithm for the vertex-restricted case.  Both
algorithms run in $O(n^3 \log n)$ time.

\end{abstract}

\newpage
\section{Introduction}
Measuring the similarity of two curves is an important problem which occurs in
many applications. A popular distance measure, that takes into account the
continuity of the curves, is the \Frechet distance.
Imagine walking forwards along both of the two curves whose similarity is to be measured.
At any point in time, the positions on the two curves have to stay within
distance~$\eps$. The minimal $\eps$ for which such a traversal is possible is the \Frechet distance.
It has been used for
simplification~\cite{ahmw-nltcs-05,bjwyz-spcdfd-08},
clustering~\cite{bbgll-dcpcs-08}, and map-matching~\cite{aerw-mpm-03,cdgnw-amm-11}.
The \Frechet distance also has applications in matching biological sequences~\cite{wz-pts-12},
analysing tracking data~\cite{bpsw-mmvtd-05,bbg-dsfm-08}, and matching coastline data~\cite{mdbh-cmpdfd-06}.

\begin{wrapfigure}[9]{r}{0.33\textwidth}
      \raggedleft
      \vspace{-1.5\baselineskip}
      \includegraphics{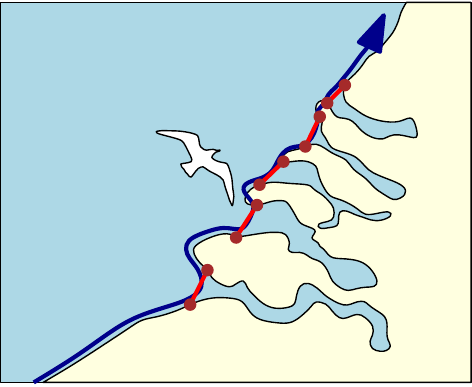}
\end{wrapfigure}
Despite its versatility, the \Frechet distance has one serious drawback: it is
a bottleneck distance. Hence it is quite sensitive to outliers, which are
frequent in real world data sets. To remedy this Driemel and
Har-Peled~\cite{dh-jydfd-11} introduced a variant of the \Frechet distance,
namely the \emph{shortcut \Frechet distance}, that allows shortcuts from and to
any point along one of the curves. The shortcut \Frechet distance is then
defined as the minimal \Frechet distance over all possible such shortcut
curves.

The shortcut \Frechet distance automatically cuts across outliers and allows us to
ignore data specific ``detours'' in one of the curves. Hence it produces
more meaningful results when dealing with real world data than the classic
\Frechet distance. Consider the following two examples. Birds are known to use
coastlines for navigation, e.g., the Atlantic flyway for migration. However,
when the coastline takes a ``detour'', like a harbor or the mouth of a river,
the bird ignores this detour, and instead follows a ``shortcut'' across. See the
example of a seagull in the figure, navigating along the coastline of Zeeland.
Using the shortcut \Frechet distance,
we can detect this similarity.
Now imagine a hiker following a pilgrims route. The hiker will occasionally
detour from the route, for breaks along the way.
In the former example, shortcuts are allowed on the coastline, in the latter on
the hiker's path.

\paragraph{Related work}
The standard \Frechet distance can be computed in time roughly quadratic in the
complexity of the input curves~\cite{ag-cfdbt-95, bbmm-fswd-12}.
Driemel and Har-Peled introduce the notion of the shortcut \Frechet distance
and describe approximation algorithms in the restricted case where shortcuts
have to start and end at input vertices~\cite{dh-jydfd-11}. In particular, they
give a $(3 + \eps)$-approximation algorithm for the vertex-restricted shortcut
\Frechet distance that runs in $O(n \log^3 n)$ time under certain input assumptions.
Specifically, they assume $c$-packedness, that is, the length of the input curves in any
ball is at most $c$ times the radius of the ball, where $c$ is a constant.
Their algorithm also yields a polynomial-time exact algorithm to compute the
vertex-restricted shortcut \Frechet distance that runs in $O(n^5 \log n)$ time and uses $O(n^4)$
space without using any input assumptions~\cite{d-raapg-13}.

The shortcut \Frechet distance can be interpreted as a partial distance
measure, that is, it maps parts of one curve to another curve. In contrast to
other partial distance measures, it is parameter-free. A different notion of a
partial \Frechet distance was developed by Buchin et al.~\cite{bbw-09}.
They propose the total length of the longest subcurves which lie
within a given \Frechet distance to each other as similarity measure.
The parts omitted are completely ignored, while in our definition
these parts are substituted by shortcuts, which have to be matched under the \Frechet
distance.

\paragraph{Results and Organization}
We study the complexity of computing the shortcut \Frechet distance.
Specifically, we show that in the general case, where shortcuts can be taken at
any point along a curve, the problem of computing the shortcut \Frechet
distance exactly is NP-hard.  This is the first hardness result for a variant
of the \Frechet distance between two polygonal curves in the plane.

Below we give an exact definition of the problem we study. In
Section~\ref{sec:nphard} we describe a reduction from SUBSET-SUM to the
decision version of the shortcut \Frechet distance and in
\secref{np:hard:correct} we prove its correctness.  The NP-hardness stems from
an exponential number of combinatorially different shortcut curves which
cause an exponential number of reachable components in the free space diagram.
We use this in our reduction together with a mechanism that controls the
sequence of free space components that may be visited.

In Section~\ref{sec:approx} we discuss polynomial-time solutions for approximation.
We give two algorithms for the decision version of
the problem, which both run in  $O(n^3 \log n)$ time.
The decision algorithms traverse the free space (as usual for \Frechet distance algorithms), and make
use of a line stabbing algorithm of Guibas et al.~\cite{ghms-91} to test
whether shortcuts are admissible.  The first algorithm uses a crucial lemma of
Driemel and Har-Peled~\cite{dh-jydfd-11} to approximate the reachable free
space and so prevents it from being fragmented.  This yields a
$3$-approximation algorithm for the decision version of the general shortcut
\Frechet distance.
The second algorithm concerns the vertex-restricted case. Here, the free space
naturally does not fragment, however, the line-stabbing algorithm helps us to
improve upon the running time of the exact decision algorithm described by
Driemel~\cite{d-raapg-13}.
We conclude with an extensive discussion of open problems in
\secref{conclusions}. In particular we
discuss some challenges in extending the decision algorithm to the computation
problem.

\paragraph{Definitions}
A \emph{curve} $\cX$ is a continuous mapping from $[0,1]$ to $\Re^2$, where $\cX(t)$ denotes the point on the curve parameterized by $t \in [0,1]$.
Given two curves $\cX$ and $\cY$ in $\Re^2$, the \emph{\Frechet
       distance} between them is
    \[
    \distFr{\cX}{\cY} = \inf_{%
       % \substack{f:[0,1] \rightarrow [0,1],\\ g:[0,1]\rightarrow
       %    [0,1] }}~
       \substack{f:[0,1] \rightarrow [0,1] }}~ \max_{\alpha \in [0,1]}
    % \distX{\cX(f(\alpha))}{\cY(g(\alpha))},
    \distX{\cX(f(\alpha))}{\cY(\alpha)},
    \]
    where $f$ is an orientation-preserving homeomorphism. %of $\cX$.
We call the line segment between two arbitrary points $\cY(\y)$ and $\cY(\y')$ on $\cY$
a \emph{shortcut} on $\cY$.
Replacing a number of subcurves of $\cY$ by the shortcuts
connecting their endpoints results in a shortcut curve of $\cY$. Thus,
a \emph{shortcut curve} is an order-preserving concatenation of non-overlapping
subcurves of $\cY$ that has straight line segments connecting the endpoints of
the subcurves.

Our input are two polygonal curves: the \emph{\targetX{}} $\cX$ and \emph{\baseY{}}~$\cY$.
The \emph{shortcut \Frechet distance} $\distSFr{\cX}{\cY}$ is now defined as the minimal \Frechet
distance between the \targetX{} $\cX$ and any shortcut curve of the \baseY{}~$\cY$.

\section{NP-hardness reduction}\label{sec:nphard}
\seclab{np:hard}
We prove that deciding if the shortcut \Frechet distance between two given
curves is smaller or equal a given value is weakly NP-hard by a reduction from
SUBSET-SUM.  We first discuss the main ideas and challenges in
\secref{np:hard:idea}, then we formally describe the reduction in
\secref{np:hard:reduction}. The correctness is proven in
\secref{np:hard:correct}.

\subsection{General idea}\seclab{np:hard:idea}

The SUBSET-SUM instance is given as a set of input values and a designated sum
value. A solution to the instance is a subset of these values that has the
specified sum.
We describe how to construct a \emph{\targetX{}} $\cX$ and a \emph{\baseY{}}
$\cY$, such that there exists a shortcut curve of $\cY$
which is in \Frechet distance $1$ to $\cX$ if and only if there
exists a solution to the SUBSET-SUM instance. We call such a shortcut curve
\emph{feasible} if it lies within \Frechet distance $1$. The construction of
the curves is split into gadgets, each of them encoding one of the input values.

\begin{figure}
  \centering
      \includegraphics{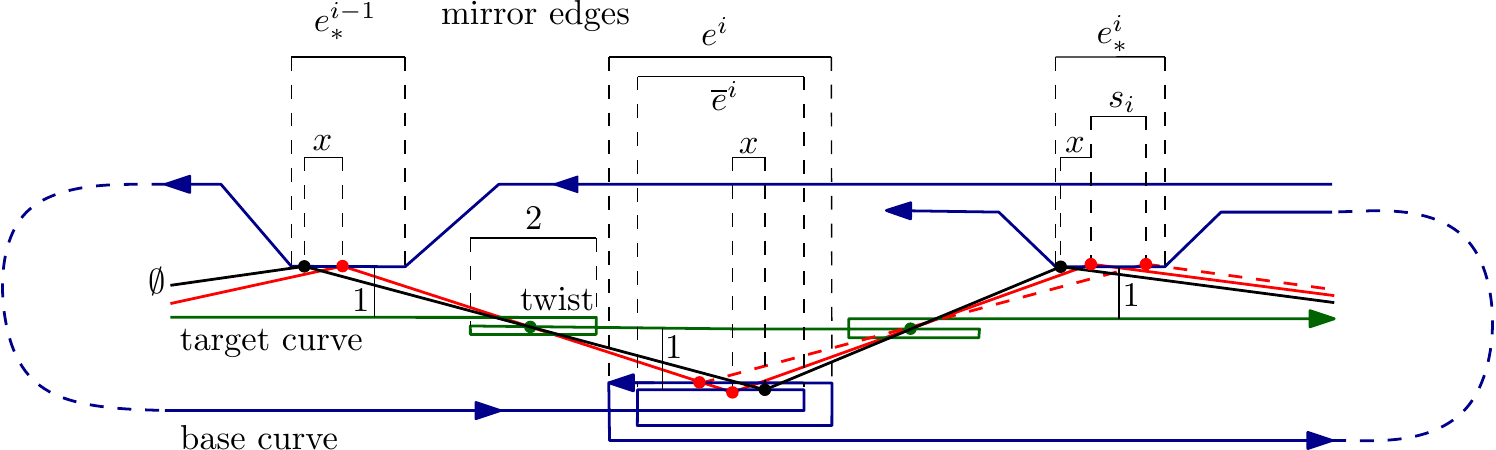} %[width=\textwidth]
      \caption{Simplistic version of the  gadget encoding one of the input
      values $s_i$. A shortcut curve entering from a position on the edge $\edge^{i-1}_{*}$ has the
choice to visit either $\edge^{i}$ or $\overline{\edge}^{i}$. Its distance on $\edge^{i}_{*}$ to the shortcut curve 
encoding the empty set ($x$ in the figure) is altered by $s_i$. The \targetX{} is distorted to show
its topology.}
      \figlab{simple:split}
\end{figure}

The idea of the reduction is as follows.
We construct the \targetX{} to lie on a horizontal line going mostly rightwards.  The \baseY{} has
several horizontal edges which lie at distance exactly $1$ to the \targetX{} and which go
leftwards. We call these edges ``mirrors'' for reasons that will become clear later.  All other
edges of the \baseY{} lie at distance greater than $1$ to the \targetX{}.  If a shortcut curve is
feasible, then any of its shortcuts must start where the previous shortcut ended. This way, the
shortcut curve ``jumps'' rightwards along the mirrors of the \baseY{} and visits each edge in
exactly one point.  We restrict the solution space of feasible shortcut curves further by letting
the \targetX{} go leftwards for a distance $2$ and then rightwards again  (see
\figref{simple:split}).  We call this a ``twist''. We place the mirrors far enough from any twist,
such that any shortcut curve which is feasible has to go through the center of each such
twist, since it has to traverse the twist region using a shortcut and this shortcut has to have
\Frechet distance at most $1$ to the twisting portion. 

The reader may picture the shortcut curve as a lightbeam, which is reflected in all directions when
it hits a mirror. In this analogy, the \targetX{} is a wall which has a hole at the center of
each twist, thus, these are the only points where light can go through. Only if we can send light
from the first vertex of the \baseY{} to the last vertex of the \baseY{}, there exists a 
feasible shortcut curve. This curve describes the path of one photon from the first to the last vertex 
of the \baseY{}. 

Using the basic mechanism of twists and mirrors, we can transport information rightwards along the
shortcut curve as follows.  Assume we have a shortcut curve that encodes the empty set. It
describes the path of a photon emitted from the first vertex of the \baseY{}. 
Assume that another photon, which took a different path, is reflected at distance $x$ along the same
mirror. If both photons also travel through the next twist center, they will hit the next
mirror at the same distance $x$ from each other.

We can offer a choice to the photon by placing two mirror edges $\edge^{i}$ and
 $\overline{\edge}^{i}$ in its line of direction (see \figref{simple:split}). 
The choice of the edge changes the position at which the photon will hit the next 
mirror edge. In particular, by placing the mirror edges carefully, we can encode 
one of the input values $s_i$ in this horizontal shift.  
By visiting a number of such gadgets, which have been threaded together, a shortcut curve
accumulates the sum of the selected subset in its distance to the curve that encodes the empty set.
We construct the terminal gadget such that only those photons that selected a subset which
sums to the designated value can see the last vertex of the \baseY{} through the last hole in the
wall.

There are two aspects of this construction we need to be careful with. First, if we want to offer a
choice of mirrors to the same photon, we cannot place both of them at distance exactly $1$ to the
\targetX{}. We will place one of them at distance $\alpha=1/2$. In this case, it may happen that a
shortcut curve visits the edge in more than one point by moving leftward on the edge before leaving
again in rightward direction. Therefore, the visiting position which encodes the current partial sum
will be only approximated.  We will scale the problem instance to prevent influence of this
approximation error on the solution.  Secondly, if two mirror edges overlap horizontally, such as
$\edge^{i}$ and $\overline{\edge}^{i}$ in \figref{simple:split}, a photon could visit both of them.
We will use more than two twists per gadget to realize the correct spacing (see \figref{main_gadget}).

\subsection{Reduction}\seclab{np:hard:reduction}

We describe how to construct the curves $\cX$ and $\cY$ from an instance of
SUBSET-SUM and how to extract a solution.

\paragraph{Input} We are given $n$ positive integers $\brc{s_1, s_2, \dots,
s_{n}}$ and a positive integer $\sigma$.  The problem is to decide whether there
exists an index set $I$, such that $\sum_{i \in I} s_i = \sigma$.  For any index
set $I$, we call $\sigma_i = \sum_{1 \leq j \leq i, j \in I} s_j$ the $i$th
\emph{partial sum} of the corresponding subset.

\paragraph{Global layout}
We describe global properties of the construction and introduce basic terminology.
Our construction consists of $n+2$ gadgets: an initialization gadget
$\gadget{0}$, a terminal gadget $\gadget{n+1}$,
and split gadgets $\gadget{i}$ for each value $s_i,$ for $i=1,\ldots,n$.
A \emph{gadget} $\gadget{i}$ consists of curves $\cX_i$ and $\cY_i$.
We concatenate these in the order of $i$ to form $\cX$ and $\cY$.
The construction of the gadgets $\gadget{1},\dots,\gadget{n}$ is incremental.
Given the endpoints of the last mirror edge of gadget $\gadget{i-1}$ and the
value $s_{i}$, we construct gadget $\gadget{i}$.
We denote with $\hLine{y}$ the horizontal line at $y$. 
The \targetX{} will be contained in $\hLine{0}$.
We call the locus of points that are within distance $1$ to the \targetX{} the
\emph{hippodrome}. 

The \baseY{} $\cY_i$ will have leftward horizontal edges
on $\hLine{-1}$, $\hLine{1}$ and $\hLine{\alpha}$, where $\alpha=\frac{1}{2}$ is
a global parameter of the construction. We call these edges \emph{mirror edges}.
The remaining edges of $\cY_i$, which are used to connect the mirror edges
to each other, are called \emph{connector edges}.
The mirror edges which will be located on $\hLine{1}$ and $\hLine{-1}$ can be
connected using curves that lie outside the hippodrome. Since those connector edges
cannot be visited by any feasible shortcut curve, their exact placement is
irrelevant. The edges connecting to mirror edges located on $\hLine{\alpha}$
and which intersect the hippodrome are placed carefully such that
no feasible shortcut curve can visit them.

The edges of the \targetX{} $\cX_i$ lie on $\hLine{0}$ running
in positive $x$-direction, except for occasional twists, which we define
as follows. A \emph{twist} centered at a point $\pnt=(p,0)$ is a subcurve
defined by the vertices $(p-1,0)$,$(p+1,0)$,$(p-1,0)$,$(p+1,0)$ which we connect
by straight line segments in this order. We call $\pnt$ the \emph{projection
center} of the twist. Let $\zeta=5$ be a global parameter of the construction,
we call the open rectangle of width $\zeta$ and height $3$ and centered at $\pnt$ a
\emph{buffer zone} of the twist. In our construction, the \baseY{} stays outside the
buffer zones. %of the twists of the \targetX{}.

Since all relevant points of the construction lie on a small set of horizontal
lines, we can slightly abuse notation by denoting the
$x$-coordinate of a point and the point itself with the same variable, albeit
using a different font.
\footnote{For example, we denote with $\splitPntA{j}{i}$ the point
that has $x$-coordinate~$\splitCoordA{j}{i}$.}
%In the figures, we omit the top index $i$ of the variables of gadget $\gadget{i}$
%to make the description less cluttered.

\paragraph{Global variables}
The construction uses four global variables.  The parameter $\alpha=\frac{1}{2}$ is
the $y$-coordinate of a horizontal mirror edge which does not lie on $H_1$ or $H_{-1}$ . 
The parameter $\beta=5$ controls the minimal horizontal distance
between mirror edges that lie between two consecutive buffer zones. The parameter
$\gamma>0$ acts as a scaling parameter to ensure (i) that the projections stay
inside the designated mirror edges and (ii) that the projections of two
different partial sums are kept disjoint despite the approximation error. How 
to choose the exact value of $\gamma$ will follow from
Lemma~\ref{lem:all-traversals}. 
The fourth parameter $\zeta=5$ defines the width of a buffer zone. It controls the 
minimum horizontal distance that a point on a mirror edge has to a projection
center.

\begin{table}\centering
    \includegraphics{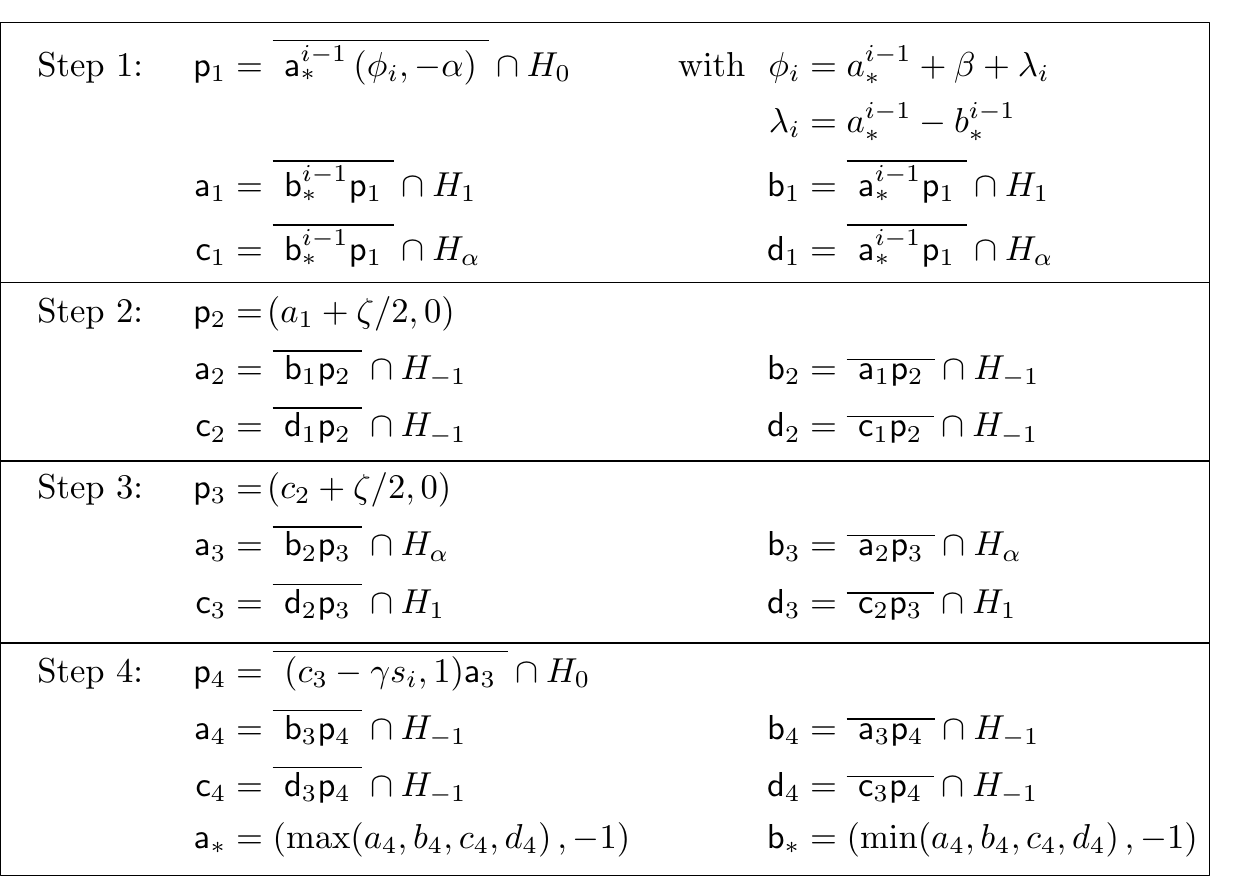}
    \caption{Construction of split gadgets $\gadget{i}$ for $1\leq i\leq n$.
       We omitted the top index $i$ of the variables.
       Thus, $\splitPntBi{*}$ stands for $\splitPntB{*}{i}$, etc.
       $\hLine{1}$, $\hLine{-1}$ and $\hLine{\alpha}$  denote the horizontal
       lines at $1,-1$ and $\alpha$ respectively.
We used $\LineY{\pntA}{\pntB}$ to denote the line through the points $\pntA$ and
$\pntB$.
}
    \label{table}
\end{table}

\begin{figure}
  \centering
  \includegraphics[width=\textwidth]{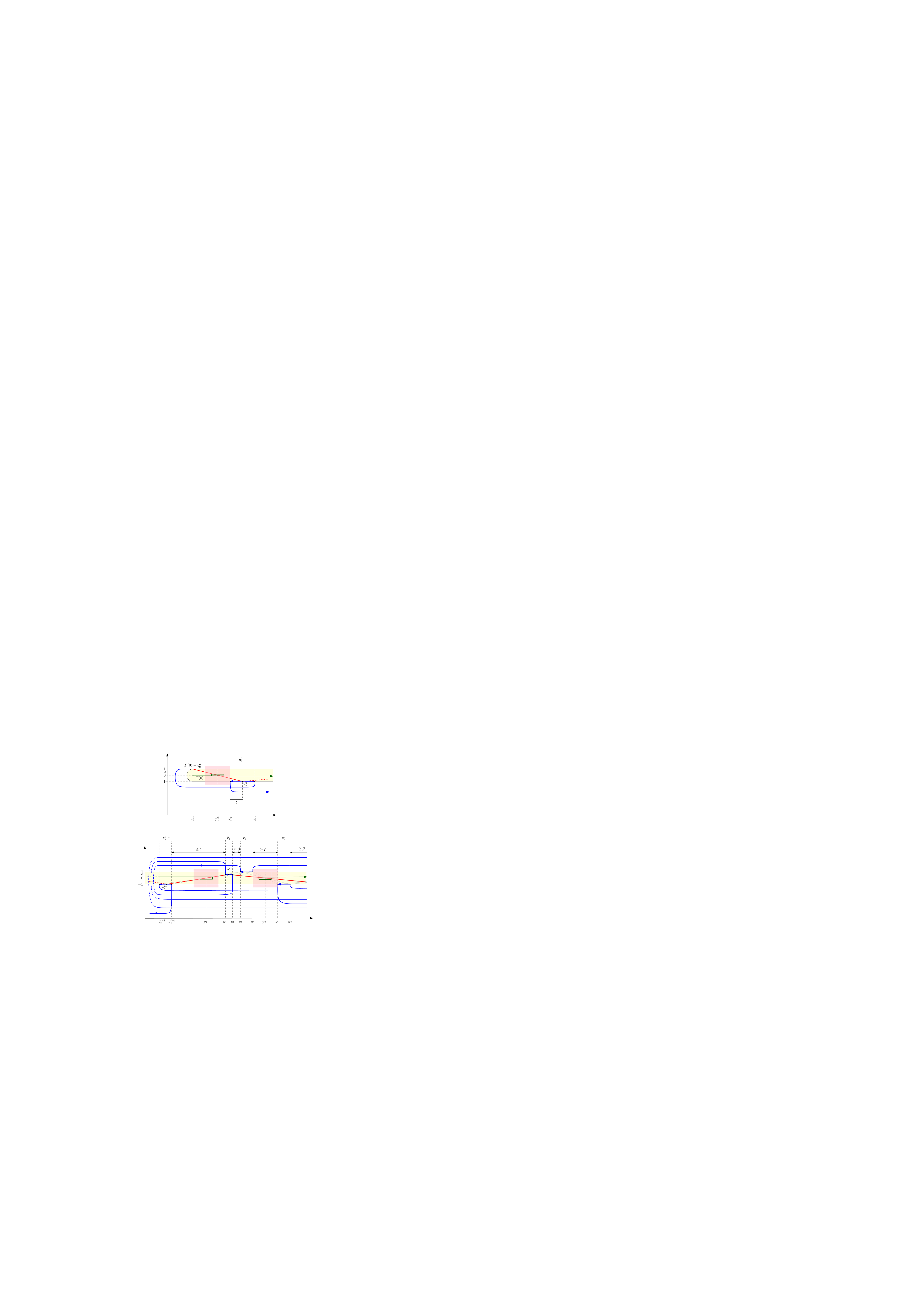}
      \caption{Top: initialization gadget $\gadget{0}$; bottom: the left part of
      split gadget $\gadget{i}$. The \targetX{} is shown in green, the \baseY{} in blue.
      An example shortcut curve is shown in red. Buffer zones and the hippodrome
      are shown as shaded regions.
      For the sake of presentation, the \targetX{} is distorted to show its
      topology, and the lengths of the mirror edges have been assumed smaller.
      The top index $i$ has been omitted from the variables.}
      \label{fig:main_gadget}
\end{figure}

\begin{figure}
  \ContinuedFloat
  \centering
  \includegraphics[width=\textwidth]{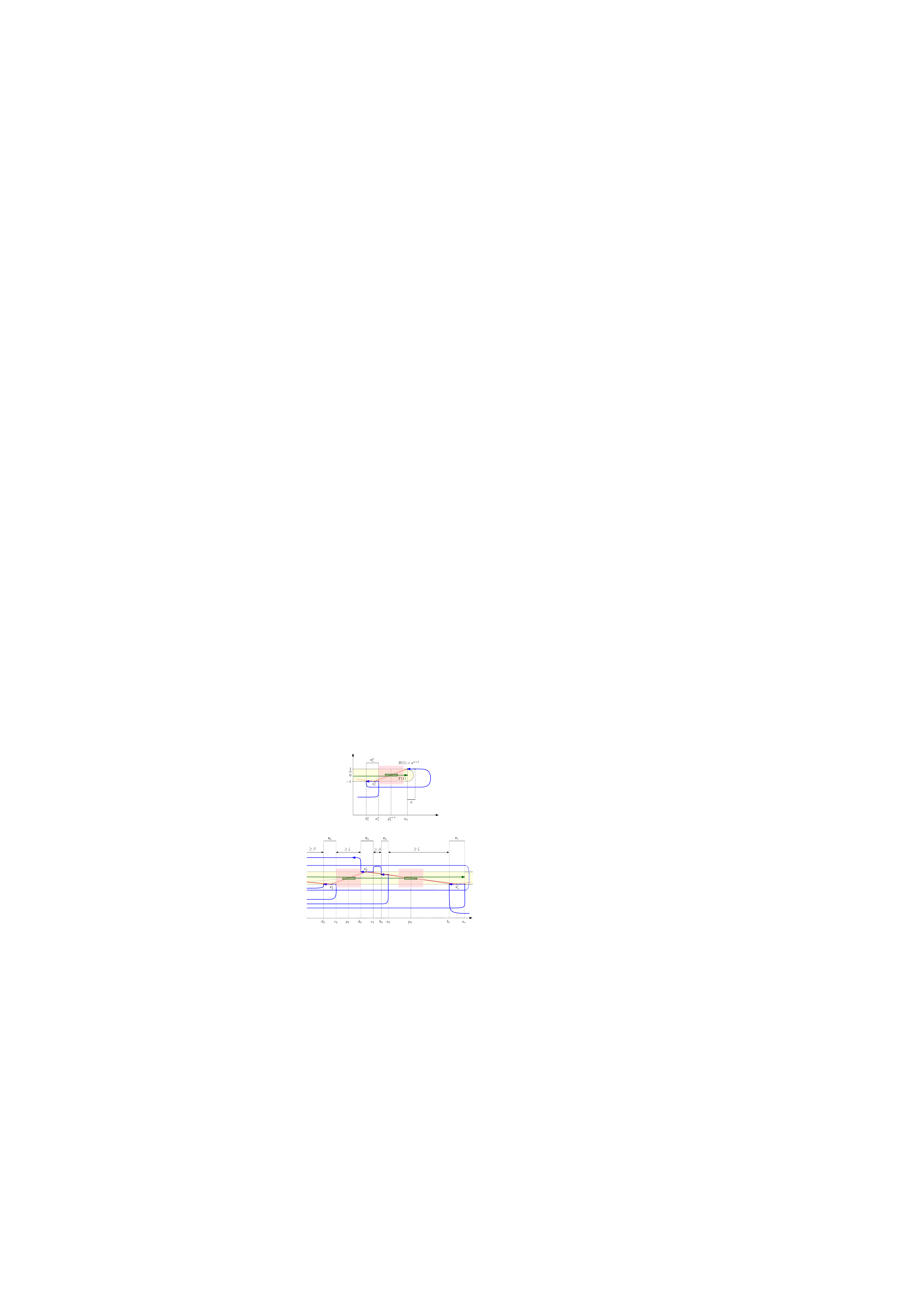}
      \caption{Top: terminal gadget $\gadget{n+1}$; bottom: the right part of
      gadget $\gadget{i}$ with example shortcut curve. \vspace{5\baselineskip} }
\end{figure}

\paragraph{Initialization gadget $(\gadget{0})$}
We let both curves start on the vertical line at $a^0_0=0$ by placing
the first vertex of $\cX{}_0$ at $(\splitCoordA{0}{0},0)$ and the first vertex of the
$\cY{}_0$ at $(\splitCoordA{0}{0},1)$.  The \baseY{} $\cY_0$ then continues to the left on $\hLine{1}$ while the
\targetX{} $\cX_0$ continues to the right on $\hLine{0}$. See
Figure~\ref{fig:main_gadget} (top left) for an illustration.
The curve $\cX{}_{0}$ has one twist centered at $\splitCoordP{1}{0} = \splitCoordA{0}{0} + \gamma +\ZoneVar/2$.
The curve $\cY{}_{0}$ has one mirror edge
$\splitEdgeYes{*}{0}=\splitPntA{*}{0}\splitPntB{*}{0}$, which we define by
setting $\splitCoordB{*}{0}=\splitCoordP{1}{0}+\ZoneVar/2$ and
$\splitCoordA{*}{0}=\splitCoordP{1}{0}+2\gamma +\ZoneVar/2$.
%Note that that $\splitPntA{0}{0}$ projects onto a point on this edge which is
%in distance $\gamma$ to both endpoints.

\paragraph{Split gadgets $(\gadget{1},\dots,\gadget{n})$}
%
%A gadget consists of a somewhat involved set of curves, the purpose will become
%clear later.
%
The overall structure is depicted in Figure~\ref{fig:main_gadget}.
The curve $\cY_i$ for $1 \leq i \leq n$ has seven mirror edges. These are
$\splitEdgeYes{j}{i}=\splitPntA{j}{i}\splitPntB{j}{i}$, and
$\splitEdgeNo{j}{i}=\splitPntC{j}{i}\splitPntD{j}{i}$, for $1 \leq j \leq 3$,
and the edge
$\splitEdgeYes{*}{i}=\splitPntA{*}{i}\splitPntB{*}{i}$.
We connect the mirror edges using additional edges to define the following order along
the \baseY{}: $\splitEdgeYes{*}{i-1}, \splitEdgeYes{1}{i}, \splitEdgeNo{1}{i},
\splitEdgeNo{2}{i}, \splitEdgeYes{2}{i}, \splitEdgeYes{3}{i},
\splitEdgeNo{3}{i}, \splitEdgeYes{*}{i}$.
The mirror edges lie on the horizontal lines $\hLine{1}$, $\hLine{-1}$ and $\hLine{\alpha}$.
We use vertical connector edges which run in positive $y$-direction and
additional connector edges which lie completely outside the hippodrome to connect
the mirror edges on $\hLine{\alpha}$.
The curve $\cX_i$ for $1\leq i \leq n$ consists of four \twist{}s centered at
the projection centers~$\splitPntP{j}{i}$ for $1 \leq j \leq 4$ which are
connected in the order of $j$ by rightward edges on $\hLine{0}$.
To choose the exact coordinates of these points, we go through several rounds of
fixing the position of the next projection center and then projecting the
endpoints of mirror edges to obtain the endpoints of the next set of mirror
edges.  The construction is defined in four steps in Table~\ref{table} and
illustrated in Figure~\ref{fig:main_gadget}~(bottom).
The intuition behind this choice of projection centers is the following.
In every step we make sure that the \baseY{} stays out of the buffer zones.
Furthermore, in Step 1 we choose the projection center far enough to the right
such that two mirror edges located between two consecutive buffer zones have
horizontal distance at least $\beta$ to each other.
In Step 4 we align the projections of the two edges $\splitEdgeYes{3}{i}$ and
$\splitEdgeNo{3}{i}$. In this alignment, the visiting position that represents
``$0$'' on $\splitEdgeYes{3}{i}$ (i.e., in its distance to $\splitPntA{3}{i}$)
and the visiting position that represents ``$s_i$'' on $\splitEdgeNo{3}{i}$
(i.e., in its distance to $\splitPntC{3}{i}$) both project to the same point
on $\splitEdgeYes{*}{i}$ (i.e., the visiting position that represents ``$s_i$''
in its distance to $\splitPntB{*}{i}$). In this way, the projections from
$\splitEdgeYes{3}{i}$ are horizontally shifted by $s_i$ (scaled by $\gamma$)
with respect to the projections from $\splitEdgeNo{3}{i}$.

\paragraph{Terminal gadget $(\gadget{n+1})$}
%Assume we have constructed the gadgets $\gadget{0}$,\dots,$\gadget{n}$ and now want
%to construct gadget $\gadget{n+1}$ from $\splitEdgeYes{n}{*}$.
The curve $\cX{}_{n+1}$ has one twist centered at $\splitCoordP{1}{n+1}=\splitCoordA{*}{n}+\ZoneVar/2$.
Let $p_\sigma=(\splitCoordB{*}{n}+\gamma(\sigma+1))$ and project the point $(p_\sigma,-1)$
through $\splitPntP{1}{n+1}$ onto $\hLine{1}$ to obtain a point $(a_\sigma,1)$.  We finish the
construction by letting both the \targetX{} $\cX_{n+1}$ and the \baseY{} $\cY_{n+1}$ end on a vertical line at $a_{\sigma}$.
The curve  $\cX_{n+1}$ ends on $\hLine{0}$ approaching from the left, while the curve
$\cY_{n+1}$ ends on $\hLine{1}$ approaching from the right.
Figure~\ref{fig:main_gadget} (top right) shows an
illustration.

\paragraph{Encoding of a subset}
Any shortcut curve of the \baseY{} encodes a subset of the
SUBSET-SUM instance.  We say the value $s_i$ is included in the encoded
subset if and only if the shortcut curve visits the edge $\splitEdgeYes{1}{i}$.
The $i$th partial sum of the encoded subset will be represented
by the point where the shortcut curve visits the edge $\splitEdgeYes{i}{*}$.
In particular, the distance of the visiting point to the endpoint of the edge
represents this value, scaled by $\gamma$ and up to a small additive error.

\section{Correctness of the reduction}
\seclab{np:hard:correct}

Now we prove that the construction has the desired behavior. That is, we prove
that any \traversal{} encodes a subset that constitutes a solution to the
SUBSET-SUM instance (Lemma~\ref{lem:all-traversals}) and
for any solution of the SUBSET-SUM instance, we can construct a \traversal{} (Lemma~\ref{lem:one-touch}).

We call a shortcut curve \emph{one-touch} if it visits any edge of the base
curve in at most one point.  Intuitively, for any feasible shortcut curve of
$\cY$, there exists a one-touch shortcut curve that ``approximates'' it.  We
first prove the correctness of the reduction for this restricted type of
shortcut curve (\lemref{one-touch}), before we turn to general shortcut curves.
In \defref{one-touch} we define a \emph{one-touch encoding}, which is a
one-touch shortcut curve that is feasible if and only if the encoded subset
constitutes a valid solution. For such curves, \lemref{any-set-sum} describes
the correspondence of the current partial sum with the visiting position on the
last edge of each gadget. It readily follows that we can construct a feasible
shortcut curve from a valid solution (\lemref{one-touch}). For the other
direction of the correctness proof we need some lemmas to testify that any
feasible shortcut curve is approximately monotone (\lemref{monotone}), has its
vertices outside the buffer zones (\lemref{zones}), and therefore has to go through all
projection centers (\lemref{projection-centers}).  We generalize
\lemref{any-set-sum} to bound the approximation error in the representation of the
current partial sum (\lemref{more-touch}).  As a result, \lemref{all-traversals}
implies that any feasible shortcut curve encodes a valid solution.

\begin{wrapfigure}[7]{r}{0.31\textwidth}
     \centering
     \vspace{-.75\baselineskip}
     \includegraphics[width=0.31\textwidth]{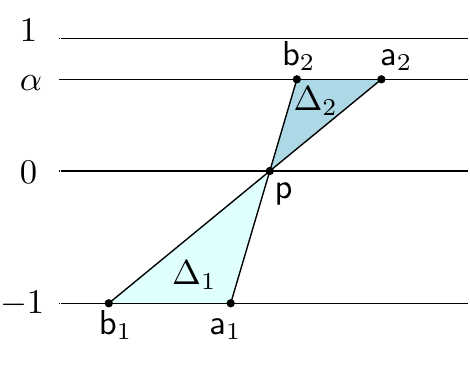}
\end{wrapfigure}

\paragraph{Distance projection}
In the correctness proof we often reason by using projections of distances between
$\hLine{1}$,$\hLine{-1}$ and $\hLine{\alpha}$.  The common argument is captured
in the following observation.

\begin{observation}[\DistanceProjection]\label{obs:distance-projection}
If $\Delta_1$ is a triangle defined by two points $\splitPntA{1}{}$ and
$\splitPntB{1}{}$ that lie on $\hLine{-1}$ and a point
$\splitPntP{}{}$ that lies on $\hLine{0}$,
and if $\Delta_2$ is a triangle defined by the points $\splitPntP{}{}$,
and the two points
$\splitPntB{2}{}=\LineY{\splitPntA{1}{}}{\splitPntP{}{}} \cap \hLine{\alpha}$
and
$\splitPntA{2}{}=\LineY{\splitPntB{1}{}}{\splitPntP{}{}} \cap \hLine{\alpha}$,
which are the projections onto $\hLine{\alpha}$,
then it holds that
$\alpha(a_1-b_1)=a_2-b_2$, where $a_1,b_1,a_2$ and $b_2$ are the respective
$x$-coordinates of the points.

\end{observation}

\subsection{Correctness for one-touch shortcut curves}

We first analyze our construction under the simplifying assumption that only
shortcut curves are allowed that are one-touch, i.e., shortcut curves can
visit the base curve in at most one point per edge. In the next section we will
build upon this analysis for the general case.

\begin{defn}[One-touch encoding]\label{def:one-touch}
Let $I$ be an index set of a subset $S'\subseteq S$.  We construct a one-touch
shortcut curve $\cY_I$ of the \baseY{} incrementally. The first two vertices
on the initial gadget are defined as follows.
We choose the first vertex of the \baseY{} $\cY(0)$ for $\splitVtx{0}{0}$, then we
project it through the first projection center $\splitPntP{1}{0}$ onto
$\splitEdgeYes{*}{0}$ to obtain $\splitVtx{*}{0}$.  Now for $i>0$,
if $i \in I$, then we project $\splitVtx{*}{i-1}$
through $\splitPntP{1}{i}$ onto $\splitEdgeYes{1}{i}$, otherwise onto
$\splitEdgeNo{1}{i}$ to obtain $\splitVtx{1}{i}$.
We continue by projecting $\splitVtx{j}{i}$ through $\splitPntP{j+1}{i}$ onto
$\cY_i$ to obtain $\splitVtx{j+1}{i}$, for $1 \leq j \leq 4$. Let
$\splitVtx{*}{i}=\splitVtx{4}{i}$.
We continue this construction throughout all gadgets in the order of $i$.
Finally, we choose $\cY(1)=(a_\sigma,1)$ as the last vertex of our shortcut curve.
Figure~\ref{fig:main_gadget} shows an example.
\end{defn}

\begin{figure}[tb]
     \centering
      \includegraphics[width=\textwidth]{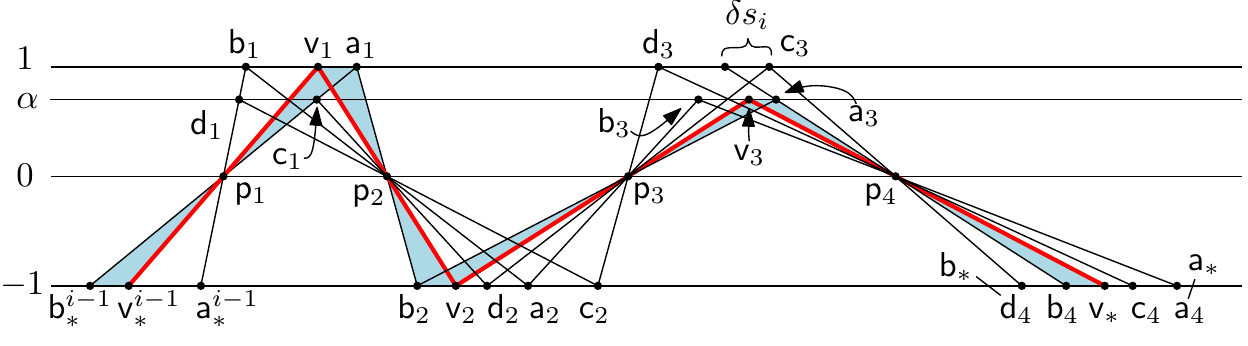}
      \caption{The path of a shortcut curve through the gadget $\gadget{i}$ in
      the case where $s_i$ is included in the selected set
      (Lemma~\ref{lem:any-set-sum}). For presentation purposes, we allowed the
      mirror edges to overlap horizontally and we omitted the top index $i$ of
      the variables.
     }
      \label{fig:one-touch-geometry}
\end{figure}

\begin{lemma}\lemlab{estar}
For any $1\leq i\leq n$, it holds that
$\splitCoordB{4}{i}=\splitCoordB{*}{i}+\gamma s_i$ and
that $\splitCoordD{4}{i}=\splitCoordB{*}{i}$.
\end{lemma}
\begin{proof}
By construction, the projection center $\splitPntP{4}{i}$ lies on a common line with
$\splitPntA{3}{i}$ and $\splitPntB{4}{i}$. Recall that we chose $\splitPntP{4}{i}$
such that this line intersects $\hLine{1}$ at the $x$-coordinate $
\splitCoordC{3}{i} - \gamma s_i$ (see Table~\ref{table} and \figref{one-touch-geometry}).
Thus, by Observation~\ref{obs:distance-projection}, it holds that
$\splitCoordB{4}{i}-\splitCoordD{4}{i}=\gamma s_i$.
Furthermore,
$\splitPntD{4}{i}$ is the point with minimum $x$-coordinate out of the projections
of $\splitPntA{3}{i}$, $\splitPntB{3}{i}$, $\splitPntC{3}{i}$, and $\splitPntD{3}{i}$
through $\splitPntP{4}{i}$ onto $\hLine{-1}$, since $\gamma s_i \geq 0$.
Since we chose $\splitPntB{*}{i}$ as such point with minimum $x$-coordinate, the
claim is implied.
\end{proof}

\begin{lemma}\label{lem:any-set-sum}
Given a shortcut curve $\cY_I$ which is a one-touch encoding
(\defref{one-touch}), let $\splitVtx{*}{i}$ be the vertex of $\cY_I$ on
$\splitEdgeYes{*}{i}$, for any $0 \leq i\leq n$. It holds for the distance
of $\splitVtx{*}{i}$ to the endpoint of this edge that
$\distX{\splitVtx{*}{i}}{\splitPntB{*}{i}} = \gamma(\sigma_i+1)$, where
$\sigma_i$ is the $i$th partial sum of the subset encoded by $\cY_I$.
\end{lemma}

\begin{proof}
We prove the claim by induction on $i$. For $i=0$, the claim is true by the construction of the
initialization gadget, since the partial sum $\sigma_0 = 0$ and
$\distX{\splitVtx{*}{0}}{\splitPntB{*}{0}} =  \gamma$.
For $i>0$,
there are two possibilities, either $i \in I$ or $i \notin I$.
Consider the case that $s_i$ is included in the set encoded by $\cY_I$.
In that case, the curve has to visit the edge~$\splitEdgeYes{1}{i}$.

By a repeated application of Observation~\ref{obs:distance-projection} we can derive that
\[
\distX{\splitVtx{*}{i-1}}{\splitPntB{*}{i-1}}
= \distX{\splitVtx{1}{i}}{\splitPntA{1}{i}}
= \distX{\splitVtx{2}{i}}{\splitPntB{2}{i}}
= \frac{\distX{\splitVtx{3}{i}}{\splitPntA{3}{i}}}{\alpha}
= \distX{\splitVtx{*}{i}}{\splitPntB{4}{i}}.
\]
Refer to \figref{one-touch-geometry} for an illustration of the geometry of the
path through the gadget. The shaded region shows the triangles that transport
the distances.
Therefore, by induction and by \lemref{estar},
\[
\distX{\splitVtx{*}{i}}{\splitPntB{*}{i}}
= \distX{\splitVtx{*}{i}}{\splitPntB{4}{i}} + \distX{\splitPntB{4}{i}}{\splitPntD{4}{i}}
= \gamma(\sigma_{i-1}+1) + \gamma s_i
= \gamma(\sigma_i+1).
\]
Thus, the claim follows for the case that $s_i$ is selected.
For the second case,
the curve has to visit the edge $\splitEdgeNo{1}{i}$.
Again, by Observation~\ref{obs:distance-projection} it holds that
\[
\distX{\splitVtx{*}{i-1}}{\splitPntB{*}{i-1}}
= \frac{\distX{\splitVtx{1}{i}}{\splitPntC{1}{i}}}{\alpha}
= \distX{\splitVtx{2}{i}}{\splitPntD{2}{i}}
= \distX{\splitVtx{3}{i}}{\splitPntC{3}{i}}
= \distX{\splitVtx{*}{i}}{\splitPntD{4}{i}}.
\]
Thus $\distX{\splitVtx{*}{i}}{\splitPntD{4}{i}} = \gamma(\sigma_{i-1}+1) =
\gamma(\sigma_i+1)$. By \lemref{estar}, $\splitPntD{4}{i}=\splitPntB{*}{i}$, thus
the claim is implied also in this case.
\end{proof}

Using the arguments from the proof of \lemref{any-set-sum} one can derive the
following corollary.

\begin{corollary}\corlab{edge-lengths}
For any $ 0 \leq i \leq n$ the length of the edge $\splitEdgeYes{*}{i}$ is equal
to $\gamma\pth{\sum_{1\leq j\leq i} s_j + 2}$.
\end{corollary}

\begin{lemma}\label{lem:zones}
If $\alpha \in [1/2,1)$ and $\beta \geq \ZoneVar/2$, then the \baseY{} does not
enter any of the buffer zones.
\end{lemma}
\begin{proof}
For the buffer zones centered at $\splitPntP{2}{i}$ and $\splitPntP{3}{i}$ for $1 \leq i
\leq n$ the claim is implied by construction. The same holds for the projection
centers of the initialization gadget and the terminal gadget.  Thus, we only
need to argue about the first and the last projection center of the intermediate
gadgets $\gadget{i}$ for $1 \leq i\leq n$.
Consider $\splitPntP{1}{i}$, by construction and since $\alpha\geq 1/2$, it holds that
\[
\splitCoordD{1}{i}-\splitCoordP{1}{i}=\splitCoordP{1}{i}-\phi_i \geq \phi_i - \splitCoordA{*}{i-1}
= \beta + \lambda_i > \beta \geq \ZoneVar/2,
\]
where $\lambda_i$ and $\phi_i$ are defined as in the construction of the gadgets. Since $\splitCoordD{1}{i}$ is
the closest $x$-coordinate of the \baseY{} to $\splitPntP{1}{i}$, the claim follows for the
first projection center of a gadget.

For the last projection center, $\splitPntP{4}{i}$, we use the fact that it lies to the right
of the point where the line through $\splitPntA{3}{i}$ and $\splitPntC{3}{i}$ passes through $\hLine{0}$.
Let the $x$-coordinate of this point be denoted $c_i$.  Now, let $\Delta_1$ be the
triangle defined by $\splitPntP{2}{i}$, $\splitPntD{1}{i}$ and $\splitPntC{1}{i}$ and
let $\Delta_2$ be the triangle defined by $\splitPntP{3}{i}$, $\splitPntA{3}{i}$ and
$\splitPntB{3}{i}$.  By the symmetry of the construction, the two triangles are
the same up to reflection at the bisector between $\splitPntP{2}{i}$ and
$\splitPntP{3}{i}$. Therefore,
\[
\splitCoordP{4}{i} - \splitCoordA{3}{i} \geq c_i-\splitCoordA{3}{i} =
\splitCoordD{1}{i}-\splitCoordP{1}{i} > \ZoneVar/2,
\]
and this implies the claim.
\end{proof}

\begin{lemma}\label{lem:monotone}
Any feasible shortcut curve is rightwards $\MonotoneConst$-monotone.
That is, if $x_1$ and $x_2$ are the $x$-coordinates of two points that
appear on the shortcut curve in that order, then $x_2 + 4 \geq x_1$.
Furthermore, it lies inside or on the boundary of the hippodrome.
\end{lemma}

\begin{proof}
Any point on the feasible shortcut curve has to lie within distance $1$ to some
point of the \targetX{}, thus the curve cannot leave the hippodrome.
As for the montonicity,
assume for the sake of contradiction, that there exist two points such that $x_2
+ 4 < x_1$. Let $\widehat{x}_1$ be the $x$-coordinate of the point on \targetX{}
matched to $x_1$ and let $\widehat{x}_2$ be the one for $x_2$. By the \Frechet
matching it follows that $\widehat{x}_2 - 1 + 4 < \widehat{x}_1 + 1$. This would
imply that the \targetX{} is not 2-monotone, which contradicts the way we
constructed it.
\end{proof}

\begin{lemma}\label{lem:projection-centers}
If $\alpha \in [1/2,1)$, $\ZoneVar>4$, and $\beta \geq \ZoneVar/2$, then a
\traversal{} passes through every buffer zone of the \targetX{} via its projection
center and furthermore it does so from left to right.
\end{lemma}

\begin{proof}
Any \traversal{} has to start at $\cY(0)$ and end at $\cY(1)$, and all of its vertices
must lie in the hippodrome or on its boundary.
By Lemma~\ref{lem:zones} the \baseY{} does not enter any of the buffer zones and
therefore  the \traversal{} has to pass through the buffer zone by using a shortcut.
If we choose the width of a buffer zone $\ZoneVar>\MonotoneConst$, then  the
only manner possible to do this while matching to the two associated vertices
of the \targetX{} in their respective order, is to go through the intersection
of their unit disks that lies at the center of the buffer zone. This is the
projection center associated with the buffer zone. By the order in which the mirror
edges are connected to form the base curve, it must do so in positive
$x$-direction and it must do so exactly once.
\end{proof}

\begin{lemma}\label{lem:beta_1}
For any $1 \leq i \leq n$ it holds that
$\splitCoordB{1}{i}-\splitCoordC{1}{i}\geq \beta$,
$\splitCoordD{2}{i}-\splitCoordA{2}{i}\geq \beta$, and
$\splitCoordB{3}{i}-\splitCoordC{3}{i}\geq \beta$.
\end{lemma}
\begin{proof}
Recall that we chose $\splitPntP{1}{i}$ by constructing the point $(\phi_i,-\alpha)$, where
$\phi_i=\splitCoordA{*}{i-1}+ \lambda_i +\beta $ and $\lambda_i =
\splitCoordA{*}{i-1}-\splitCoordB{*}{i-1}$. The construction is such that
$(\phi_i,-\alpha)$, $\splitPntP{1}{i}$ and $\splitPntA{*}{i-1}$ lie on the same
line.
Consider the point $(r_i,-\alpha)$ that lies on the line through $\splitPntB{*}{i-1}$ and
$\splitPntP{1}{i}$. We have that the triangle $\Delta_1$ defined by $(r_i,-1)$,
$\splitPntP{1}{i}$ and $\splitPntA{*}{i-1}$ is the same up to rotation as the
triangle $\Delta_2$ defined by $(\splitCoordC{1}{i},1)$,
$\splitPntP{1}{i}$ and $\splitPntB{1}{i}$. Refer to Figure~\ref{fig:beta} for an illustration. %of the geometry.
By Observation~\ref{obs:distance-projection},  % Thus,
\[
\splitCoordB{1}{i}-\splitCoordC{1}{i} = r_i - \splitCoordA{*}{i-1} = \beta +
\lambda_i - (\phi_i - r_i) \geq \beta,
\]
since $(\phi_i-r_i) = \alpha\lambda_i$. This proves the first part of the claim.
Now it readily follows that also $\splitCoordD{2}{i}-\splitCoordA{2}{i}\geq
\beta$. Indeed, it follows by Observation~\ref{obs:distance-projection} that
$q_i - \splitCoordA{2}{i} = \splitCoordB{1}{i} - \splitCoordC{1}{i}$, where
$q_i$ is the $x$-coordinate of the projection of $(\splitCoordC{1}{i},1)$
through $\splitPntP{2}{i}$ onto $\hLine{-1}$, and this projection
lies between $\splitPntD{2}{i}$ and $\splitPntA{2}{i}$. Again, refer to
Figure~\ref{fig:beta} and in particular to triangles $\Delta_1'$ and $\Delta_2'$.

\begin{figure}[b]
     \centering
      \includegraphics[width=\textwidth]{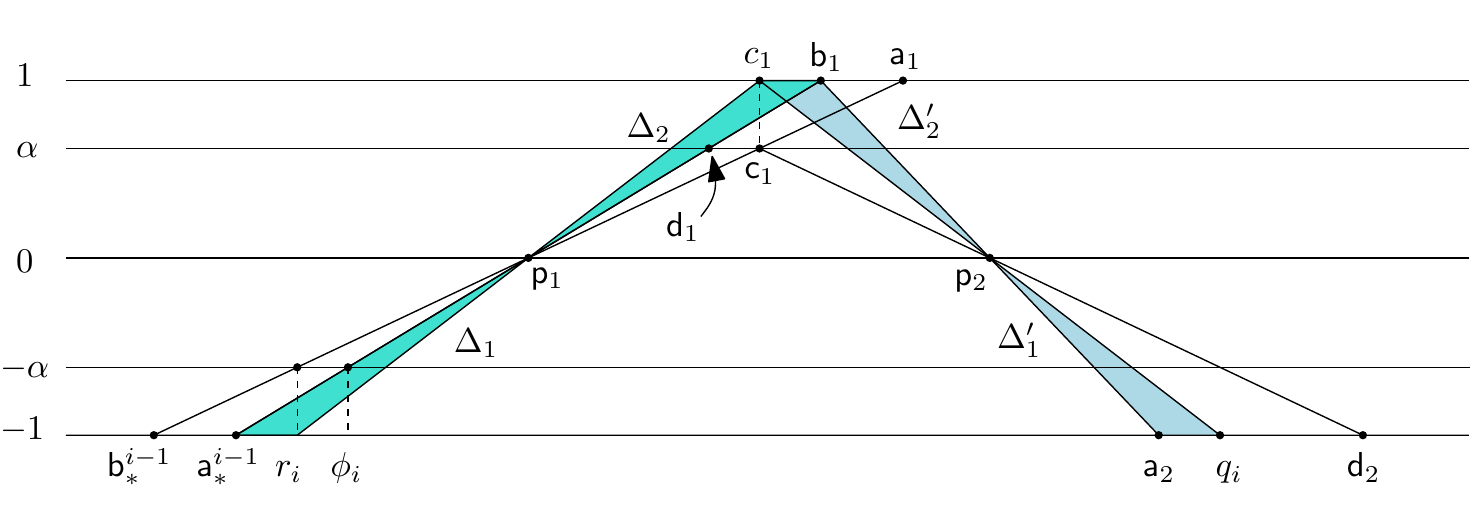}
      \caption{The geometry used in the proof of Lemma~\ref{lem:beta_1}.
       We omitted the top index $i$ of the variables.
      }
      \label{fig:beta}
\end{figure}

The claim $\splitCoordB{3}{i}-\splitCoordC{3}{i}\geq \beta$ follows from the
symmetry of the middle part of the gadget.  Consider the triangle $\Delta_3$
defined by $(\splitCoordC{1}{i},1)$, $\splitPntP{2}{i}$ and
$(\splitCoordB{1}{i},1)$ and the triangle $\Delta_4$ defined by $(\splitCoordC{3}{i},1)$,
$\splitPntP{3}{i}$ and $(\splitCoordB{3}{i},1)$. By construction $\Delta_3$ is a reflected
version of $\Delta_4$, where the axis of reflection is the bisector of the two
projection centers. Thus, by the above argument we have that
$\splitCoordB{3}{i}-\splitCoordC{3}{i} = \splitCoordB{1}{i}-\splitCoordC{1}{i} \geq \beta$.
\end{proof}

\begin{lemma}\lemlab{either-or}
If $\alpha \in [1/2,1)$, $\ZoneVar>\MonotoneConst$ and if $\beta>\MonotoneConst$, then a
\traversal{} that is one-touch visits either $\splitEdgeYes{j}{i}$ or
$\splitEdgeNo{j}{i}$ for any $1 \leq i \leq n$ and $1 \leq j \leq 3$.
Furthermore, it visits all edges $\splitEdgeYes{*}{i}$ for $0 \leq i \leq n$.
\end{lemma}
\begin{proof}
By \lemref{monotone}, any feasible shortcut curve is $\MonotoneConst$-monotone. Furthermore,
it starts at $\cY(0)$ and ends at $\cY(1)$ and by \lemref{projection-centers},
it goes through all projection centers of the target curve from left to right.
We first want to argue that it visits at least one mirror edge between two
projection centers, i.e., that it cannot ``skip'' such a mirror edge by matching
to two twists in one shortcut. Such a shortcut would have to lie on $\hLine{0}$,
since it has to go through the two corresponding projection centers lying on
$\hLine{0}$. By construction, the only possible endpoints of such a shortcut lie
on the connector edges that connect to mirror edges on $\hLine{\alpha}$.
Assume such a shortcut could be taken by a shortcut curve starting from $\cY(0)$.
Then, there must be a connector edge which intersects a line from a point on a
mirror edge through a projection center. In particular, since the curve has to
go through all projection centers, one or more of the following must be true for
some $1\leq i \leq n$:
(i) there exists a line through $\splitPntP{1}{i}$ intersecting a mirror edge
$\splitEdgeYes{*}{i-1}$ and a connector edge of
$\splitEdgeNo{1}{i}$, or (ii) there exists  a line through $\splitPntP{3}{i}$
intersecting a mirror edge $\splitEdgeYes{2}{i}$ or $\splitEdgeNo{2}{i}$ and a
connector edge of $\splitEdgeYes{3}{i}$. However, this was
prevented by the careful placement of these connector edges.

It remains to prove that the shortcut curve cannot visit both
$\splitEdgeYes{j}{i}$  and $\splitEdgeNo{j}{i}$ for any $i$ and $j$, and
therefore visits at most one mirror edge between two projection centers. First
of all, the shortcut curve has to lie inside or on the boundary of the hippodrome and is $4$-monotone
(\lemref{monotone}). At the same time, we constructed the gadget such that
the mirror edges between two consecutive
projection centers have distance at least $\beta$ by \lemref{beta_1} and that
the left mirror edge comes after the right mirror edge along their order of
$\cY$.  Since we chose $\beta>4$, the shortcut curve cannot visit both mirror edges.
\end{proof}

Putting the above lemmas together implies the correctness of the reduction for
shortcut curves that are one-touch, i.e., which visit every edge in at most one
point.

\begin{lemma} \lemlab{one-touch}
If $\alpha \in [1/2,1)$, $\ZoneVar>\MonotoneConst$ and if $\beta>\MonotoneConst$, then
for any feasible one-touch shortcut curve $\cY_{\Diamond}$, it holds that the
subset encoded by $\cY_{\Diamond}$ sums to $\sigma$. Furthermore, for any subset
of $s$ that sums to $\sigma$, there exists a feasible one-touch shortcut curve that
encodes it.
\end{lemma}

\begin{proof}
\lemref{either-or} and \lemref{projection-centers} imply that  $\cY_{\Diamond}$ must be a
one-touch encoding as defined in \defref{one-touch} if it is feasible.
By \lemref{any-set-sum}, the second last vertex of $\cY_{\Diamond}$ is the point
on the edge $\splitEdgeYes{*}{n}$, which is in distance $\gamma(\sigma_{\Diamond}+1)$ to
$\splitPntB{*}{n}$, where $\sigma_{\Diamond}$ is the sum encoded by
the subset selected by $\cY_{\Diamond}$.
The last vertex of $\cY_{\Diamond}$ is equal to $\cY(1)$, which we
placed in distance $\gamma(\sigma+1)$ to the projection of $\splitPntB{*}{n}$
through $\splitPntP{1}{n+1}$. Thus, if and ony if $\sigma_{\Diamond}=\sigma$,
then the
last shortcut of $\cY_{\Diamond}$ passes through the last projection center of
the \targetX{}.  It follows that if $\sigma_{\Diamond} \neq \sigma$, then
$\cY_{\Diamond}$ cannot be feasible.
For the second part of the claim, we construct a one-touch encoding as defined in
\defref{one-touch}. By the above analysis, it will be feasible if the subset
sums to $\sigma$, since the curve visits every edge of $\cY$ in at most one
point and in between uses shortcuts which pass through every buffer zone from left to
right and via the buffer zone's projection center.
\end{proof}

\subsection{Size of the construction}

We prove that the construction has polynomial size.

\begin{lemma}\lemlab{size}
The curves can be constructed in $O(n)$ time.  Furthermore, if we choose $\alpha = 1/2, \beta=5$,
$\ZoneVar=5$, and $\gamma = 25n $, then the size of the coordinates used is in $O(\log n +
\log(\sum_{i=0}^{n} s_i))$.
\end{lemma}
\begin{proof}
Each of the constructed gadgets uses a constant number of vertices. Since we construct $n+2$
gadgets, the overall number of vertices used is in $O(n)$. The curves can be constructed using a
single iteration from left to right, therefore they can be constructed in $O(n)$ time.

Secondly, we can bound the size of the coordinates as follows.
We claim that
\begin{equation}
\splitCoordA{\sigma}{}
\in O\pth{n^2 \sum_{i=0}^{n} s_i}
\end{equation}

Using basic geometry, we can bound the horizontal length of an individual split gadget as follows.
Let $A_i=\splitCoordP{1}{i}-\splitCoordA{*}{i-1}$.
Let $\lambda_{i}=\splitCoordA{*}{i-1}-\splitCoordB{*}{i-1}$ as defined as in Table~\ref{table} (see also Figure~\ref{fig:main_gadget}).
%
%\maike{how about this slightly more condensed form of the equation?}
%\begin{align*}
% \splitCoordA{*}{i}-\splitCoordA{*}{i-1}
%&= (\splitCoordA{*}{i}-\splitCoordP{4}{i}) + (\splitCoordP{4}{i}-\splitCoordD{3}{i}) + (\splitCoordD{3}{i}-\splitCoordC{2}{i}) + (\splitCoordC{2}{i}-\splitCoordP{2}{i}) + (\splitCoordP{2}{i}-\splitCoordB{1}{i}) + (\splitCoordB{1}{i} - \splitCoordP{1}{i}) + (\splitCoordP{1}{i}-\splitCoordA{*}{i-1})\\
%&= (\splitCoordA{*}{i}-\splitCoordP{4}{i}) + (\splitCoordP{4}{i}-\splitCoordD{3}{i}) + \ZoneVar + (\splitCoordC{2}{i}-\splitCoordP{2}{i}) + (\lambda_i +\ZoneVar/2) + A_i + A_i\\
%&= (\splitCoordA{*}{i}-\splitCoordP{4}{i}) + (\splitCoordP{4}{i}-\splitCoordD{3}{i}) + (\splitCoordC{2}{i}-\splitCoordP{2}{i}) + \lambda_i + 3\ZoneVar/2 + 2A_i\\
%&\leq (\splitCoordA{*}{i}-\splitCoordP{4}{i}) + (\splitCoordP{4}{i}-\splitCoordD{3}{i}) + \pth{\frac{\splitCoordP{2}{i}-\splitCoordD{1}{i}}{\alpha}} + \lambda_i + 3\ZoneVar/2 + 2A_i\\
%&= (\splitCoordA{*}{i}-\splitCoordP{4}{i}) + (\splitCoordP{4}{i}-\splitCoordD{3}{i}) + \pth{\frac{A_i + \lambda_i + \ZoneVar/2}{\alpha}} + \lambda_i + 3\ZoneVar/2 + 2A_i\\
%&\leq (\splitCoordA{*}{i}-\splitCoordP{4}{i}) + (\splitCoordP{4}{i}-\splitCoordD{3}{i}) + 3A_i + 2\lambda_i + 2\ZoneVar\\
%&\leq 2\pth{\frac{\splitCoordA{3}{i}-\splitCoordD{3}{i} + \gamma s_i}{1-\alpha}}+ 3A_i + 2\lambda_i + 2\ZoneVar\\
%&\leq 2\pth{\frac{A_i +\lambda_i  + \gamma s_i}{1-\alpha}}+ 3A_i + 2\lambda_i + 2\ZoneVar\\
%&= 4A_i + 3\lambda_i + \gamma s_i + 2\ZoneVar
%\end{align*}
%

\begin{align*}
 \splitCoordA{*}{i}-\splitCoordA{*}{i-1}
&= (\splitCoordA{*}{i}-\splitCoordB{1}{i}) + (\splitCoordB{1}{i} - \splitCoordP{1}{i}) +
(\splitCoordP{1}{i}-\splitCoordA{*}{i-1})\\
&= (\splitCoordA{*}{i}-\splitCoordB{1}{i}) + 2A_i\\
&= (\splitCoordA{*}{i}-\splitCoordP{2}{i} )+ (\splitCoordP{2}{i} -\splitCoordB{1}{i})+ 2A_i\\
&= (\splitCoordA{*}{i}-\splitCoordP{2}{i} )+ \lambda_i +\ZoneVar/2 + 2A_i\\
&= (\splitCoordA{*}{i}- \splitCoordC{2}{i})+(\splitCoordC{2}{i} -  \splitCoordP{2}{i}) + \lambda_i +\ZoneVar/2 + 2A_i\\
&\leq (\splitCoordA{*}{i}- \splitCoordC{2}{i})+\pth{\frac{\splitCoordP{2}{i}-
\splitCoordD{1}{i}}{\alpha}} + \lambda_i +\ZoneVar/2 + 2A_i\\
&= (\splitCoordA{*}{i}- \splitCoordC{2}{i})+\pth{\frac{A_i + \lambda_i + \ZoneVar/2}{\alpha}} + \lambda_i +\ZoneVar/2 + 2A_i\\
&\leq (\splitCoordA{*}{i}- \splitCoordC{2}{i})+ 4A_i + 3\lambda_i + 2\ZoneVar\\
&= (\splitCoordA{*}{i}- \splitCoordD{3}{i})+(\splitCoordD{3}{i} -\splitCoordC{2}{i})+ 4A_i +
3\lambda_i + 2\ZoneVar\\
&= (\splitCoordA{*}{i}- \splitCoordD{3}{i})+ 4A_i + 3\lambda_i + 3\ZoneVar\\
\displaybreak[1]
&= (\splitCoordA{*}{i}-\splitCoordP{4}{i})+(\splitCoordP{4}{i}-\splitCoordD{3}{i})+ 4A_i + 3\lambda_i + 3\ZoneVar\\
&\leq 2\pth{\frac{\splitCoordA{3}{i}-\splitCoordD{3}{i} + \gamma s_i}{1-\alpha}}+ 4A_i + 3\lambda_i + 3\ZoneVar\\
&\leq 2\pth{\frac{A_i +\lambda_i  + \gamma s_i}{1-\alpha}}+ 4A_i + 3\lambda_i + 3\ZoneVar\\
&= 8A_i + 7\lambda_i + 4\gamma s_i + 3\ZoneVar
\end{align*}

By the construction of the first projection center in Table~\ref{table}, it holds that
\begin{equation*}
A_i \leq \frac{\beta+\lambda_i}{1-\alpha} =  2(\beta+\lambda_i)
\end{equation*}

Putting everything together, we get
\begin{align*}
\splitCoordA{\sigma}{}
&= (\splitCoordA{\sigma}{}-\splitCoordA{*}{n})
+ \sum_{i=1}^{n} ( \splitCoordA{*}{i}-\splitCoordA{*}{i-1})
+ \splitCoordA{*}{0}\\
&\leq (2\lambda_{n+1} + \ZoneVar)
+ \sum_{i=1}^{n} ( \splitCoordA{*}{i}-\splitCoordA{*}{i-1})
+  (\ZoneVar + 3\gamma)\\
&\leq  2 \lambda_{n+1} + \ZoneVar
+ \sum_{i=1}^{n}\pth{ 16\pth{ \beta+\lambda_i }
+ 7\lambda_i  + 4\gamma s_i + 3\ZoneVar } + \ZoneVar + 3\gamma\\
&\leq
23\sum_{i=1}^{n+1} \lambda_{i}
+4\gamma \sum_{i=1}^{n} s_i
+16n\beta + 3(n+1)\ZoneVar
+ 3\gamma
\end{align*}

By construction of the initialization gadget, it holds that $\lambda_1=2\gamma$. Together with
 \corref{edge-lengths} this implies that
\[\lambda_i \leq \gamma\pth{\sum_{i=1}^{n} s_j + 2}\] for $1\leq i \leq n+1$.
Since we chose $\gamma=25n$ and the remaining global variables constant, the claim is implied.
\end{proof}

\subsection{Correctness for general shortcut curves}

Next, we generalize Lemma \ref{lem:any-set-sum} to bound the incremental
approximation error of the visiting positions on the last edge of each gadget
for general shortcut curves, that is, assuming shortcut curves are not necessarily one-touch.

\begin{lemma}\lemlab{more-touch}
Choose $\alpha \in [1/2,1)$, $\ZoneVar>\MonotoneConst$ and
$\beta>\MonotoneConst$.
Given a feasible shortcut curve $\scCurve{\cY}$, let $\splitVtx{*}{i}$ be any point of
$\scCurve{\cY}$ on $\splitEdgeYes{*}{i}$ and let $\splitCoordV{*}{i}$ denote its $x$-coordinate.
For any $0 \leq i\leq n$ let $\sigma_i$ denote the $i$th partial sum of the
subset encoded by $\scCurve{\cY}$.  If we choose $\gamma>\eps_i$, then it holds
that
\[
\splitCoordB{*}{i} + \gamma_i - \eps_i
\leq \splitCoordV{*}{i}
%\leq \splitCoordA{*}{i} - (\gamma - \eps_i)
\leq \splitCoordB{*}{i} + \gamma_i + \eps_i
\]
where $\gamma_i=\gamma(\sigma_i+1)$ and
$\eps_i=\frac{8i+4}{\alpha}$.
\end{lemma}
\newcommand{\minError}{\ensuremath{\gamma_{\min}}}
\newcommand{\maxError}{\ensuremath{\gamma_{\max}}}
\newcommand{\VisitInterval}[2]{\ensuremath{\Interval_{#1}^{#2}}}
\begin{proof}
We prove the claim by induction on $i$. For $i=0$ the claim follows by the
construction of the initialization gadget. Indeed, the curve $\scCurve{\cY}$ has
to start at $\cY(0)=\splitPntA{0}{0}$ and by \lemref{projection-centers} it has
to pass through both $\splitPntP{1}{0}$ and $\splitPntP{1}{1}$.  Since the three
points $\splitPntA{0}{0}, \splitPntP{1}{0},$ and $\splitPntP{1}{1}$ do not lie
on a common line, there must be an edge of the \baseY{} in between the two
projection centers visited by $\scCurve{\cY}$.
By \lemref{monotone}, the shortcut curve cannot leave the
hippodrome. However, the only edge available in the
hippodrome is $\splitEdgeYes{*}{0}$.  By construction, the only possible
shortcut to this edge ends at the center of the edge in distance $\gamma$ to
${\splitPntB{*}{0}}$. Since the mirror edge runs leftwards, the only other
points that can be visited by $\scCurve{\cY}$ lie therefore in this direction.
However, by \lemref{monotone}, $\scCurve{\cY}$ is rightwards
$\MonotoneConst$-monotone.
It follows that
\[
%\distX{\splitCoordV{*}{0}}{\splitPntB{*}{0}} \in [\gamma - 4,\gamma].
 \splitCoordB{*}{0} + \gamma - \MonotoneConst
 \leq \splitCoordV{*}{0}
 %\leq \splitCoordA{*}{i} - \gamma.
 \leq \splitCoordB{*}{i} + \gamma.
\]
Since $\eps_0 = 4/\alpha > 4$ and $\sigma_0=0$, this implies
the claim for $i=0$.

For $i>0$, the curve $\scCurve{\cY}$ entering gadget $\gadget{i}$ from the edge
$\splitEdgeYes{*}{i-1}$ has to pass through the first buffer zone via the projection
center $\splitCoordP{1}{i}$.
By induction,
\[
\splitCoordB{*}{i-1} + \minError
~\leq~
\splitCoordV{*}{i-1}
~\leq~
%\splitCoordC{*}{i-1} - \maxError,
\splitCoordB{*}{i-1} + \maxError,
\]
where
$\minError = \gamma_{i-1} - \eps_{i-1}$
and
$\maxError = \gamma_{i-1} + \eps_{i-1}$
denote the minimal and maximal distances of the visiting position to the left
endpoint on the edge $\splitEdgeYes{*}{i-1}$.
Since $\gamma > \eps_i = \eps_{i-1} + 8/\alpha$, and
$\sigma_{i-1} \geq 0$, it follows that
\begin{equation}\label{right}
\minError = \gamma(\sigma_{i-1} + 1) - \eps_{i-1} \geq \gamma - \eps_{i-1} > 8/\alpha.
\end{equation}
Furthermore, by \corref{edge-lengths},
\begin{equation}\label{left}
\maxError
%= \gamma(\sigma_{i-1} + 1) + \eps_{i-1}
\leq  \gamma\pth{\sum_{1\leq j\leq i-1} s_j + 1} + \eps_{i-1}
\leq  \pth{\lambda_i - \gamma} + \eps_{i-1}
\leq \lambda_i -  8/\alpha,
\end{equation}
where $\lambda_i = \splitCoordA{*}{i-1} - \splitCoordB{*}{i-1}$
is the length of edge $\splitEdgeYes{*}{i-1}$.
Thus, $\splitVtx{*}{i-1}$ lies at distance at least $8/\alpha$ from each endpoint
of $\splitEdgeYes{*}{i-1}$.
Therefore, the only two edges of the \baseY{} which intersect the line
$\LineY{\splitVtx{*}{i-1}}{\splitPntP{1}{i}}$ within the hippodrome are
$\splitEdgeYes{1}{i}$ and $\splitEdgeNo{1}{i}$. Note that also the vertical
connector edges at $\splitEdgeNo{1}{i}$ do not intersect any such line within the
hippodrome.

Now, there are two cases, either the shortcut ends on $\splitEdgeYes{1}{i}$  or
on $\splitEdgeNo{1}{i}$. Assume the latter case.
By \obsref{distance-projection} the $x$-coordinate of the endpoint of the shortcut
lies in the interval
\[
\pbrc{
\splitCoordC{1}{i} - \alpha \maxError
~,~
\splitCoordC{1}{i} - \alpha \minError
}.
\]
By the same observation, the length of the edge $\splitEdgeNo{1}{i}$ is equal to
$\alpha\lambda_i$. Thus, the endpoint of the shortcut lies inside the edge.

We now argue that the shortcut curve has to leave the edge by using a shortcut,
i.e., the shortcut curve cannot ``walk'' out of the edge by using a subcurve of
$\cY$.
The mirror edges are oriented leftwards.  Since the shortcut curve has to be
rightwards $\MonotoneConst$-monotone (\lemref{monotone}), it can only walk by a
distance $\MonotoneConst$ on each such edge.
Let $\VisitInterval{j}{i}$ denote the range of $x$-coordinates of
$\scCurve{\cY}$ on $\splitEdgeNo{j}{i}$. By the above,
\[
\VisitInterval{1}{i}
\subseteq
\pbrc{
\splitCoordC{1}{i} - \alpha \maxError - \MonotoneConst
~,~
\splitCoordC{1}{i} - \alpha \minError
}
\]
Thus, by Eq.~(\ref{left}) and Eq.~(\ref{right}) and since
$\splitCoordC{1}{i}-\splitCoordD{1}{i}=\alpha\lambda_i$,
it holds that
$\VisitInterval{1}{i} \subseteq [\splitCoordD{1}{i}, \splitCoordC{1}{i}] $,
i.e., the shortcut curve must leave the edge $\splitEdgeNo{1}{i}$ by using a
shortcut. A shortcut to $\splitEdgeYes{1}{i}$ would violate the order along
the base curve $\cY$. Since the shortcut curve is rightwards $\MonotoneConst$-monotone
(\lemref{monotone}) and must pass a through a buffer zone via its projection center
(\lemref{projection-centers}),
the only way to leave the edge is to take a shortcut through $\splitPntP{2}{i}$.
The only edge intersecting a line through  $\splitPntP{2}{i}$ and a point on
$\splitEdgeNo{1}{i}$ is $\splitEdgeNo{2}{i}$. Thus, $\splitEdgeNo{2}{i}$ must be the next edge
visited.
Now we can again use \obsref{distance-projection} to project the set of visiting
points onto the next edge and use the fact that  the shortcut curve can only
walk rightwards and only by a distance at most $4$ on a mirror edge to derive that
\[
\VisitInterval{2}{i} \subseteq
\pbrc{
\splitCoordD{2}{i} + \minError - \MonotoneConst
~,~
\splitCoordD{2}{i} + (\maxError + \MonotoneConst/\alpha)
}
\subseteq [\splitCoordD{2}{i}, \splitCoordC{2}{i}].
\]
By repeated application of the above arguments, we obtain
that $\splitEdgeNo{3}{i}$ is visited within
\[
\VisitInterval{3}{i} \subseteq
\pbrc{
\splitCoordC{3}{i} - (\maxError + \MonotoneConst/\alpha) - \MonotoneConst
~,~
\splitCoordC{3}{i} - (\minError - \MonotoneConst)
}
\subseteq [\splitCoordD{3}{i}, \splitCoordC{3}{i}],
\]
and that $\splitEdgeNo{*}{i}$ is visited within
\[
\VisitInterval{*}{i} \subseteq
\pbrc{
\splitCoordD{4}{i} + (\minError - \MonotoneConst) - \MonotoneConst
~,~
%\splitCoordC{4}{i} - (\maxError - 4/\alpha - 4)
\splitCoordD{4}{i} + (\maxError + \MonotoneConst/\alpha + \MonotoneConst)
}
\subseteq [\splitCoordD{4}{i}, \splitCoordC{4}{i}]
\subseteq [\splitCoordB{*}{i}, \splitCoordA{*}{i}]
\]
For each visited edge, it follows by Eq.~(\ref{left}) and Eq.~(\ref{right}) that the shortcut curve visits the edge in the interior.

Now, since the shortcut curve did not visit $\splitEdgeYes{1}{i}$, the input value
$s_i$ is not included in the selected subset, therefore
$\gamma_i=\gamma_{i-1}$. Using the interval $\VisitInterval{*}{i}$ derived above,
and the fact that $\splitCoordD{4}{i}=\splitCoordB{*}{i}$ (\lemref{estar})
it follows that
\[
\splitCoordV{*}{i}
~\geq~
\splitCoordD{4}{i} + \minError - 8
~\geq~
\splitCoordD{4}{i} + (\gamma_{i-1} - \eps_{i-1}) - 8
\geq
%\splitCoordB{*}{i} + \gamma_{i} - \eps_{i} - 8 + 8/\alpha
%~\geq~
\splitCoordB{*}{i} + \gamma_{i} - \eps_{i},
\]
and similarly,
\[
\splitCoordV{*}{i}
\leq
\splitCoordD{4}{i} + \maxError + 8/\alpha
\leq
\splitCoordD{4}{i} + (\gamma_{i-1} + \eps_{i-1}) + 8/\alpha
%\splitCoordB{*}{i} + \maxError + 8/\alpha
\leq
\splitCoordB{*}{i} + \gamma_{i} + \eps_{i}
\]
Thus, the claim follows in the case that $\scCurve{\cY}$ visits $\splitEdgeNo{1}{i}$.

The case that $\scCurve{\cY}$ visits $\splitEdgeYes{1}{i}$ can be proven along
the same lines.  However, now $s_i$ is included in the selected subset, and
therefore $\gamma_i=\gamma_{i-1}+ \gamma s_i$.
Using the arguments above we can derive
\[
\VisitInterval{*}{i}\subseteq
\pbrc{
\splitCoordB{4}{i} + \minError - 8/\alpha
~,~
\splitCoordB{4}{i} + \maxError + 8/\alpha
}
\]
By \lemref{estar}, $\splitCoordB{4}{i} = \splitCoordB{*}{i} + \gamma s_i$.
(Note that the same argument was used in \lemref{any-set-sum}).
Thus, analogous to the above
\[
\splitCoordV{*}{i}
\geq
\splitCoordB{4}{i} + \gamma_{i-1} - \eps_{i}
\geq
\splitCoordB{*}{i} + \gamma s_i + \gamma_{i-1} - \eps_{i}
\geq
\splitCoordB{*}{i} + \gamma_{i} - \eps_{i}
\]
and similarly,
\[
\splitCoordV{*}{i}
\leq
\splitCoordB{4}{i} + \gamma_{i-1} + \eps_{i}
\leq
\splitCoordB{*}{i} + \gamma s_i + \gamma_{i-1} + \eps_{i}
\leq
\splitCoordB{*}{i} + \gamma_{i} + \eps_{i}
\]
Therefore the claim is implied also in this case.
\end{proof}

\begin{lemma}\label{lem:all-traversals}
If we choose $\alpha \in [1/2,1), \beta > \MonotoneConst$,
$\ZoneVar> \MonotoneConst$, and $\gamma \geq 25n $, then
any \traversal{} $\scCurve{\cY}$ encodes a subset of $\brc{s_1,\dots,s_n}$  that sums to $\sigma$.
\end{lemma}

\begin{proof}
Since $\scCurve{\cY}$ is feasible, it must be that it visits
$\splitEdgeYes{*}{n}$ at distance $\gamma(\sigma+1)$ to $\splitPntB{*}{n}$, since
this is the only point to connect via a shortcut through the last projection
center to the endpoint of $\cY$ and by \lemref{projection-centers} all
projection centers have to be visited.  So let
$\splitCoordV{*}{n}=\splitCoordB{*}{n}+\gamma(\sigma+1)$ be the
$x$-coordinate of this visiting point (the starting point of the last shortcut),
and let $\sigma_n$ be the sum of the subset encoded by $\scCurve{\cY}$.
\lemref{more-touch} implies that
\[
\splitCoordB{*}{n} + \gamma(\sigma_n+1) - \eps_n
\leq \splitCoordB{*}{n}+ \gamma(\sigma+1)
\leq \splitCoordB{*}{n} + \gamma(\sigma_n+1) + \eps_n,
\]
since $\eps_n = \frac{8n+4}{\alpha} < 25n = \gamma$.
Therefore,
\[
\sigma_n  - \eps_n/\gamma \leq \sigma \leq \sigma_n + \eps_n/\gamma
\]
For our choice of parameters $\eps_n/\gamma<1$. Thus, it must be that
$\sigma=\sigma_n$, since both values are integers.

\end{proof}

\subsection{Main result}

Now, together with
\lemref{one-touch} and the fact that the reduction is polynomial (\lemref{size}),
\lemref{all-traversals} implies the NP-hardness of the problem.

\begin{theorem}\label{thm:main}
The problem of deciding whether the shortcut \Frechet distance between two given curves is less or
equal a given distance is NP-hard.
\end{theorem}

\section{Algorithms}\label{sec:approx}
We give two $O(n^3 \log n)$ time algorithms for deciding the shortcut \Frechet distance.
One is a $3$-approximation algorithm for the general case, and one an exact
algorithm for the vertex-restricted case.
Both algorithms traverse the free space as usual, using a line stabbing algorithm by Guibas
et~al.~\cite{ghms-91} to test the admissability of shortcuts.

The approximation algorithm for the general case uses a crucial lemma of Driemel and
Har-Peled~\cite{dh-jydfd-11} to approximate the reachable free space and prevent it from
fragmenting.
The exact algorithm for the vertex-restricted case uses a similar lemma for efficiently
testing all possible shortcuts. In this case, the free space naturally does not fragment.

First we discuss relevant preliminaries, in particular tunnels in the free space diagram and ordered line-stabbing.

\subsection{Preliminaries}\label{sec:app:prelim}
\paragraph{Free space Diagram}%
Let $\cX, \cY$ be two polygonal curves parameterized over $[0,1]$.
The standard way to compute the \Frechet distance uses the
$\delta$-\emph{free space} of $\cX$ and $\cY$, which is a
subset of the joint parametric space of $\cX$ and $\cY$, defined as
\[
\FullFDleq{\delta}(\cX, \cY) = \brc{ (\x,\y) \in [0,1]^2 \sep{
      \distX{\cX(\x)}{\cY(\y)} \leq \delta}}.
\]
From now on, we will simply write $\FullFDleq{\delta}$.
The square $[0,1]^2$, which represents the joint parametric space,
can be broken into a (not necessarily uniform) grid called the \emph{free space
diagram}, where a vertical line corresponds to a vertex of $\cX$ and a horizontal line
corresponds to a vertex of $\cY$.

Every pair of segments of $\cX$ and $\cY$ define a \emph{cell}
in this grid.  Let $\CellXY{i}{j}$ denote the cell that corresponds to the
$i$\th edge of $\cX$ and the $j$\th edge of $\cY$. The cell
$\CellXY{i}{j}$ is located in the $i$\th column and $j$\th row of this
grid.
It is known that the free space, for a fixed $\delta$, inside such a
cell $\CellXY{i}{j}$ (i.e., $\FullFDleq{\delta} \cap
\CellXY{i}{j}$) is convex~\cite{ag-cfdbt-95}.
We will denote it with $\FCellXY{\delta}{i}{j}$.

Furthermore, the \Frechet distance between two given curves is less
or equal to $\delta$ if and only if there exists a monotone path in the free space that
starts in the lower left corner $(0,0)$ and ends in the upper right corner
$(1,1)$ of the free space diagram~\cite{ag-cfdbt-95}.
For the shortcut \Frechet distance, we
need to also allow shortcuts.
This is captured in the concept of tunnels in free space.
A shortcut segment
$\ScutCrv{\cY}{\yPnt}{\yPntQ}$ and the subcurve
$\SubCrv{\cX}{\xPnt}{\xPntQ}$ it is being matched to, correspond
in the parametric space to a
segment $\pnt\pntQ \subseteq [0,1]^2$, called a \emph{tunnel{}} and
denoted by $\xtunnel{\pnt}{\pntQ}$, where $\pnt = \pth{\xPnt, \yPnt}$
and $\pntQ = \pth{\xPntQ, \yPntQ}$.  We require $\xPnt \leq \xPntQ$
and $\yPnt \leq \yPntQ$ for monotonicity.   We call the
\Frechet distance of the shortcut segment to the subcurve the
\emph{price} of this tunnel and denote it with
$\scPrice{\pnt}{\pntQ} =
\distFr{\SubCrv{\cX}{\xPnt}{\xPntQ}}{\ScutCrv{\cY}{\yPnt}{\yPntQ}}$.
A tunnel $\xtunnel{\pnt}{\pntQ}$ is \emph{feasible} for $\delta$ if
$\pnt,\pntQ \in \FullFDleq{\delta}({\cX},{\cY})$.

Now, we define the \emph{reachable free space}, as follows
\[
 \ReachFDleqI{\delta}\pth{\cX,\cY} =
\brc{  (\xPnt, \yPnt) \in [0,1]^2 \sep{
      \distSFr{\SubCrv{\cX}{0}{\xPnt}}{\SubCrv{\cY}{0}{\yPnt }}
      \leq \delta}}.
\]
From now on, we will simply write $\ReachFDleqI{\delta}$.
This is the set of points that have an $(x,y)$-monotone path from
$(0,0)$ that stays inside the free space and otherwise uses
tunnels. We will denote the reachable space inside a cell (i.e.,
$\ReachFDleqI{\delta} \cap \CellXY{i}{j}$) with
$\RCellXY{\delta}{i}{j}$.

Finally, we will use the following well-known fact, that the \Frechet
distance between two line segments is the maximum of the distance of the
endpoints.
\begin{observation}\label{obs:convex}\obslab{f:r:segments}
    Given segments $\mathsf{a}\mathsf{b}$ and $\mathsf{c}\mathsf{d}$,
    it holds $\distFr{\mathsf{a}\mathsf{b}}{\mathsf{c}\mathsf{d}} =
    \max ( \distX{\mathsf{c}}{\mathsf{a}}$, $\distX{\mathsf{d}}{\mathsf{b}})$.
\end{observation}

\paragraph{Monotonicity of tunnel prices}
In order to prevent the reachable space from being fragmented, as it is the case
with the exact problem we showed to be NP-hard, we will approximate it. For this,
we will use the following lemma from \cite{dh-jydfd-11}.

\begin{lemma}[\cite{dh-jydfd-11}]\label{lem:monotonicity}\lemlab{monotone:shortcut}
       Given a value $\delta > 0$ and two curves $\subcX{1}$ and
       $\subcX{2}$, such that $\subcX{2}$ is a subcurve of
       $\subcX{1}$, and given two line segments $\subsegY{1}$ and
       $\subsegY{2}$, such that $\distFr{\subcX{1}}{\subsegY{1}} \leq
       \delta$ and the start (resp.\ end) point of $\subcX{2}$ is in
       distance $\delta$ to the start (resp.\ end) point of
       ${\subsegY{2}}$, then $\distFr{\subcX{2}}{\subsegY{2}} \leq
       3 \delta$.
\end{lemma}

\paragraph{Horizontal, vertical and diagonal tunnels}
We can distinguish three types of tunnels.  We call a tunnel that stays within a
column of the grid, a \emph{vertical tunnel}. Likewise, a tunnel that stays
within a row is called a \emph{horizontal tunnel}. Tunnels that span across
rows and columns are \emph{diagonal tunnels}.  Note that vertical tunnels that
are feasible for a value of $\delta$ also have a price at most $\delta$
by \obsref{f:r:segments}.  Furthermore, the shortcut which corresponds to a
horizontal tunnel lies within an edge of the input curve. Thus, shortcutting the
curve does not have any effect in this case and we can safely ignore
such horizontal tunnels.

\paragraph{Ordered line stabbing}
Guibas et~al.~\cite{ghms-91} study the problem of stabbing an ordered set
of unit radius disks with a line. In particular, one of the problems studied is the
following.  Given a series of unit radius disks  $D_1,\dots,D_n$, does there
exist a directed line $\ell$, with $n$ points $\pnt_i$, which lie along $\ell$
in the order of $i$, such that $\pnt_i \in D_i$?  As was already noted
by Guibas~et~al., their techniques can
be applied to decide whether there exists a line segment that
lies within \Frechet distance one to a given polygonal curve $X$. Simply center
the disks at the vertices of $X$ in their order along the curve and the
relationship follows from the fact that the \Frechet distance between two
line segments is the maximum of the distances of their endpoints.

The algorithm described by Guibas et~al. maintains a so called
\emph{line-stabbing wedge} that contains all points $\pnt$, such that there
is a line through $\pnt$ that visits the first $i$ disks before visiting~$\pnt$.
The algorithm runs in $O(n \log n)$ time.
We will use this algorithm,
to compute all tunnels of price at most $\delta$ starting from a particular point in the
parametric space and ending in a particular cell.

\subsection{Approximate decision algorithm}
\label{sec:app:alg}
\seclab{approx:algo}

\newcommand{\AlgorithmI}[1]{{{\texttt{\bf{#1}}}}}
\newcommand{\Algorithm}[1]{{\AlgorithmI{#1}\index{#1@{\AlgorithmI{#1}}}}}
\newcommand{\tunnelDiagonal}{\Algorithm{diagonalTunnel}\xspace}
\newcommand{\tunnelVertical}{\Algorithm{verticalTunnel}\xspace}
\newcommand{\DeciderFrCont}{\Algorithm{Decider}\xspace}
\newcommand{\FigWidth}{0.9\textwidth}

We describe an approximate decision algorithm for the \asymmetric{} \unrestricted{}
shortcut \Frechet distance. Given a value of $\delta$ and two polygonal curves
$\cX$ and $\cY$ in $\Re^2$
of total complexity $n=n_1+n_2$, the algorithm outputs either (i)
''$\distoSFr{\cX}{\cY} \leq 3\delta$``, or (ii) ''$\distoSFr{\cX}{\cY} >
\delta$``. The algorithm runs in $O(n^3\log n)$ time and $O(n)$ space.
We first discuss the challenges and then give a sketch of the algorithm.

\subsubsection{Challenge and ideas}

The standard way to solve the decision problem for the \Frechet distance and
its variants is to search for monotone paths in the free space diagram. In the
case of the shortcut \Frechet distance, this path can now use tunnels in the
free space diagram, which correspond to shortcuts on $\cY$ which are matched to
subcurves of $\cX$.  In the general version of the shortcut \Frechet distance,
the tunnels can now start and end anywhere inside the free space cells, while
in the \vrestricted{} case they are constrained to the grid of the parametric
space. In order to extend the algorithm by Driemel and Har-Peled
\cite{dh-jydfd-11}  to this case, we need a new method to compute the space
which is reachable within a free space cell.

We use the concept of line-stabbing to compute all tunnels $\xtunnel{\pnt}{\pntQ}$
of price at most $\delta$ starting from a particular point $\pnt$ in the
parametric space and ending in a particular cell.  By intersecting the
line-stabbing wedge with the edge of $\cY$ that corresponds to the cell, we
obtain a horizontal strip which represents the set of such tunnel
endpoints $\pntQ$. See \figref{line-stabbing-wedge} for an illustration.

\begin{figure}[tb]
     \centering
      \includegraphics{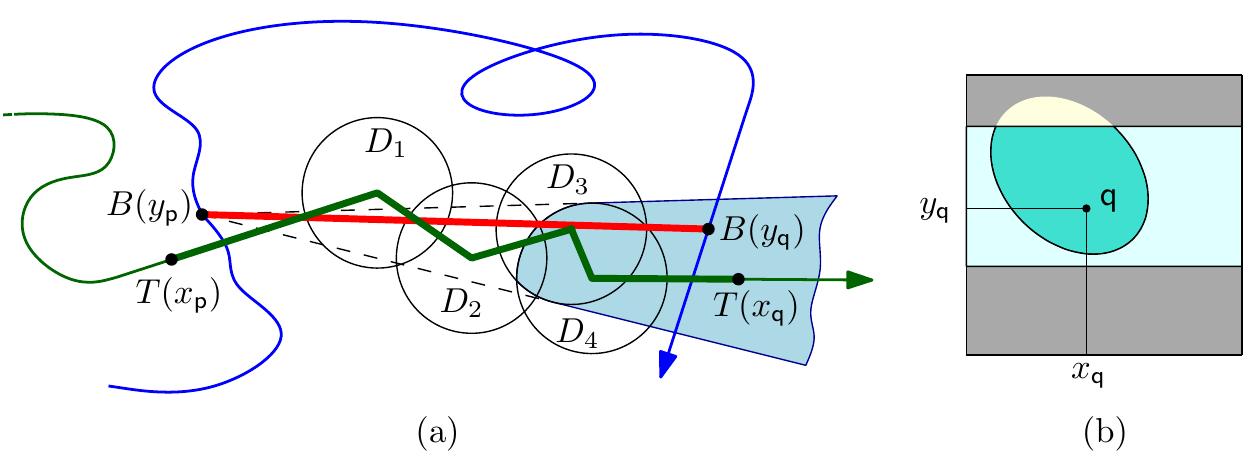}
      \caption{(a) Example of a tunnel $\xtunnel{\pnt}{\pntQ}$ computed by the
      \tunnelDiagonal{} procedure. The shaded area
      shows the line-stabbing wedge. (b) The free space cell which contains the
      endpoint of the tunnel.}
      \figlab{line-stabbing-wedge}
\end{figure}

The second challenge is that the reachable free space can fragment into
exponentially many of such horizontal strips.  However, we can exploit the
monotonicity of the tunnel prices to approximate the reachable free space as
done in the algorithm by Driemel and Har-Peled. In this approximation scheme, the
combinatorial complexity of the reachable space is constant per cell. Thus, the
overall complexity is bounded by $O(n^2)$.  Our algorithm takes $O(n \log n)$
time per cell, resulting in $O(n^3 \log n)$ time overall.

Driemel and Har-Peled make certain assumptions on the input curves and achieve
a near-linear running time. Instead of traversing $O(n^2)$ cells, they
considered only those cells that intersect the free space on their boundary. To
compute these cells, a data structure of de Berg and Streppel
\cite{bs-arsbsp-06} was used. Unfortunately, this method does not immediately
extend to our case, since we also need to consider cells that intersect the
free space in their interior only. Therefore, our algorithm traverses the entire free
space diagram yielding a much simpler, yet slower algorithm, without making
assumptions on the input curves.

\subsubsection{Sketch}
\seclab{continuous:algo:sketch}
We traverse the free space as usual to compute the reachable free space.
In each cell, in addition to the reachable free space from neighboring cells,
we also compute the free space reachable by a shortest tunnel at price $3\delta$.
For this, we store (or find) the right-most point in the free space below and
to the right of the current cell, i.e., the point that will give the shortest
tunnel.
We modify the line-stabbing algorithm of Guibas et al. to compute the free
space reachable by a tunnel from this point (see \tunnelDiagonal{} below).
Now, by the lemma of Driemel and
Har-Peled, we know that any point not reachable by a shortest tunnel at price
$3\delta$ is also not reachable by a longer tunnel at price $\delta$.
Because we compute reachability by only one tunnel, the free space fragments
only in a constant number of pieces.
In this way, we obtain a $3$-approximation to the decision version of the problem.

\subsubsection{The tunnel procedures}
\seclab{diagonal:tunnel:test}

The \tunnelDiagonal{} procedure receives as input a cell $\CellXY{i}{i}$ and a point
$\pnt \in \FullFDleq{\delta}$, such that $\pnt$ lies in the
lower left quadrant of the lower left corner of the cell $\CellXY{i}{i}$ and a parameter
$\delta > 0$.
The output will be a set of points $\PntSet \subseteq \CellXY{i}{i}$, such that
 $\scPrice{\pnt}{\pntQ} \leq \delta$ if and only if $\pntQ \in \PntSet$.
We use the line-stabbing algorithm mentioned above with minor modifications.  Let
$\xPnt=x_{-}$ be the $x$-coordinate of $\pnt$ and let $x_{+}$ be the
$x$-coordinate of some point in the interior of $\CellXY{i}{i}$.  Let
$D_1,\dots,D_k$ be the disks of radius $\delta$ centered at the vertices of
$\cX$ that are spanned by the subcurve $\SubCrv{\cX}{x_{-}}{x_{+}}$.  There are
two cases, either $\cY(\yPnt)$ is contained in $D_i$ for all $1 \leq i \leq k$,
or there exists some $i$, such that $\cY(\yPnt)$ lies outside of $D_i$. In the
first case, we return
$\PntSet = \CellXY{i}{i} \cap \FullFDleq{\delta}$.
In the second case we initialize the line-stabbing wedge of \cite{ghms-91}
with the tangents of $\cY(\yPnt)$ to the disk $D_i$, where $i$ is the
smallest index such that $\cY(\yPnt) \notin D_i$. We then proceed with the
algorithm as written by handling the disks $D_{i+1},\dots,D_k$.
Finally, we intersect the line-stabbing wedge with the edge of  \cY{} that
corresponds to $\CellXY{i}{i}$.
Refer to Figure~\ref{fig:line-stabbing-wedge} for an illustration.
This yields a horizontal slab of points that
lie in $\CellXY{i}{i}$ which we then intersect with the $\delta$-free space
and return as our set $\PntSet$.

The \tunnelVertical procedure
\seclab{vertical:tunnel:test}
receives as input a cell $\CellXY{i}{i}$ and a point $\pnt$  which lies
below this cell in the same column and a parameter $\delta \geq 0$. Let
$H_{\pnt}$ be the closed halfplane which lies to the right of the vertical line
through $\pnt$. The procedure returns the intersection of $H_\pnt$ with the
$\delta$-free space in $\CellXY{i}{i}$.

\newcommand{\Arraytype}{}
\newcommand{\currReach}{\ensuremath{\Arraytype{}\Array}}
\newcommand{\prevReach}{\ensuremath{\Arraytype{}\overline{\Array}}}
\newcommand{\currGatesMin}{\ensuremath{\Arraytype{}g^{\ell}}}
\newcommand{\prevGatesMin}{\ensuremath{\Arraytype{}\overline{g}^{\ell}}}
\newcommand{\currGatesMax}{\ensuremath{\Arraytype{}g^{r}}}
\newcommand{\prevGatesMax}{\ensuremath{\Arraytype{}\overline{g}^{r}}}

\begin{figure}[t]
    \fbox{\begin{minipage}{\FigWidth}%
           \DeciderFrCont{}$(\cX, \cY, \delta)$
           \begin{algorithmic}[1]
	       \STATE Assert that $\frVal{ 0,0} =
               \distX{\cX(0)}{\cY(0)} \leq \delta$ and $\frVal{ 1,1}
               \leq \delta$

               \STATE Let \currReach, \prevReach, \currGatesMin,
	       \prevGatesMin, \currGatesMax, and \prevGatesMax{} be arrays of
	       size $n_1$

	       \FOR{$j=1,\dots, n_2$}
		    \STATE Update $\prevReach \leftarrow \currReach$,
		    $\prevGatesMin \leftarrow \currGatesMin$, and
		    $\prevGatesMax \leftarrow \currGatesMax$

		    \FOR{$i=1,\dots, n_1$}

		       \IF {$i=1$ \AND $j=1$}
		           \STATE Let $\PntSet_{i,j}= \FCellXY{\delta}{i}{j}$
		       \ELSE
		       \STATE Retrieve $\RVCellXY{i-1}{j}$ and $\RHCellXY{i}{j-1}$ from
			      $\currReach[i-1]$ and $\prevReach[i]$

		       \STATE Step 1:
		              Compute $\PntSet^{1}_{i,j}$
			      from $\RVCellXY{i-1}{j}$ and $\RHCellXY{i}{j-1}$

		       \STATE Step 2:
			      Let $\PntSet^2_{i,j}=
		              \tunnelVertical\pth{\prevGatesMin[i],\CellXY{i}{j}, \delta}$.
		
		       \STATE Step 3:
		              Let $\PntSet^3_{i,j}=
			      \tunnelDiagonal\pth{\prevGatesMax[i-1],\CellXY{i}{j}, 3\delta}$.
	
		       \STATE Let $\PntSet_{i,j}= Q\pth{\PntSet^{1}_{i,j} \cup
		              \PntSet^{2}_{i,j} \cup \PntSet^{3}_{i,j}} \cap
		              \FCellXY{\delta}{i}{j}$
		       \ENDIF

		       \IF {$\PntSet_{i,j}\neq \emptyset$ }

		       \STATE Update $\currGatesMin[i]$ and $\currGatesMax[i]$
			       using the gates of $\PntSet_{i,j}$

		       \STATE Compute $\RVCellXY{i}{j}$ and $\RHCellXY{i}{j}$
		       from $\PntSet_{i,j}$ and store them in $\currReach[i]$
		       \ELSE
		       \STATE Update $\currGatesMax[i]$ using $\currGatesMax[i-1]$

		       \ENDIF

	            \ENDFOR

	       \ENDFOR
	
	       \IF{$(1,1) \in \currReach[n_1]$}

	       \STATE Return ``$\distoSFr{\cX}{\cY} \leq 3\delta$''

	       \ELSE

	       \STATE Return ``$\distoSFr{\cX}{\cY} > \delta$''

	       \ENDIF

           \end{algorithmic}
       \end{minipage}}%
    \caption{The decision procedure \DeciderFrCont{} for the shortcut
       \Frechet distance. }
    \figlab{decider:infty:continuous}
\end{figure}

\subsubsection{The decision algorithm}
\seclab{continuous:decider}
The algorithm is layed out in \figref{decider:infty:continuous}.
We traverse the free space diagram in a row-by-row order from bottom to
top and from left to right. For every cell, we compute a set of reachable points
$\PntSet_{i,j} \subseteq \CellXY{i}{j}$, such that
\[
\RCellXY{\delta}{i}{j}\pth{\cX,\cY}
\subseteq \PntSet_{i,j}
\subseteq \RCellXY{3\delta}{i}{j}\pth{\cX,\cY}.
\]
Thus, the set of computed points approximates the reachable free space.
From $\PntSet_{i,j}$ we compute reachability intervals
$\RVCellXY{i}{j}$ and $\RHCellXY{i}{j}$, which we define as the intersections of
$\PntSet_{i,j}$ with the top and right cell boundary.
Furthermore we compute the \emph{gates} of $\PntSet_{i,j}$, which we define
as the two points of the set with minimum and maximum $x$-coordinates.
(In \cite{dh-jydfd-11}, where tunnels were confined to the
horizontal edges of the grid, gates were defined as the extremal points
of the reachability intervals.) We keep this information for the
cells in the current and previous row in one-dimensional arrays by the index
$i$.
We use three arrays $\currReach$, $\currGatesMin$  and $\currGatesMax$ to
write the information of the current row and three arrays
$\prevReach$, $\prevGatesMin$  and $\prevGatesMax$ to store
the information from the previous row. Here,
$\currReach$ and $\prevReach$ are used to store the reachability
intervals, and $\currGatesMin$, $\prevGatesMin$, $\prevGatesMax$ $\currGatesMax$
are used to store extremal points (i.e., gates) of the computed reachable space.
In particular, $\currGatesMin[i]$ and $\prevGatesMin[i]$ store the leftmost
reachable point (i.e., gate) discovered so far that lies inside column $i$ and
in $\currGatesMax[i]$ and $\prevGatesMax[i]$ we maintain the rightmost
reachable point (i.e., gate) discovered so far that lies to the left of column $i+1$.
During the traversal, we can update this information in constant time per cell
using the gates of $\PntSet_{i,j}$, and the gates stored in
$\currGatesMax[i-1]$, $\currGatesMax[i]$ and $\currGatesMin[i]$.

\begin{figure}[t]
     \centering
      \includegraphics{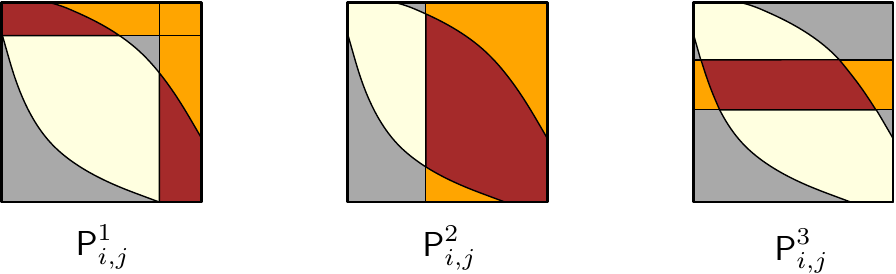}
      \caption{Examples of the approximate reachable free space in one cell:
      reachable by lower or left boundary (left), by a vertical tunnel
      (center), or a diagonal tunnel (right). (This figure is not referenced)}
      \label{fig:reach-one-cell}
\end{figure}

We handle a cell $\CellXY{i}{j}$ in three steps.  We first compute the set of
points $\PntSet^{1}_{i,j}$ in this cell that are reachable by a monotone path
via $\RVCellXY{i}{j-1}$ or $\RHCellXY{i-1}{j}$.  Since these reachability
intervals have been computed in previous steps, they can be retrieved from
$\currReach[i-1]$ and $\prevReach[i]$. More specifically, to compute
$\PntSet^{1}_{i,j}$, we take the closed halfplane above the horizontal line at
the lower endpoint of $\RVCellXY{i}{j-1}$ and intersect it with the
$\delta$-free space inside the cell, which we can compute ad-hoc from the two
corresponding edges. Similarly, we take the closed halfplane to the right of the
left endpoint of $\RHCellXY{i-1}{j}$ and intersect it with the $\delta$-free
space. The union of those two sets is $\PntSet^1_{i,j}$.
In a second step, we compute the set of points $\PntSet^{2}_{i,j}$ in $\CellXY{i}{j}$ that
are reachable by a vertical tunnel from below.  For this, we retrieve the
leftmost reachable point in the current column by probing $\prevGatesMin[i]$.
Assume there exists such a point and denote it by $\pnt_2$. We invoke
\tunnelVertical$\pth{\pnt_2, \CellXY{i}{j}, \delta}$ and let
$\PntSet^{2}_{i,j}$ be the output of this procedure.  In the third step, we
compute the set of points $\PntSet^{3}_{i,j}$ in $\CellXY{i}{j}$ that are
reachable by a diagonal tunnel. For this, we retrieve the rightmost reachable
point in the cells that are spanned by the lower left quadrant of the lower left
corner of $\CellXY{i}{j}$. This point is stored in $\prevGatesMax[i-1]$. Let
this point be $\pnt_3$, if it exists.  We invoke \tunnelDiagonal$\pth{\pnt_3,\CellXY{i}{j},
3\delta}$ and let $\PntSet^{3}_{i,j}$ be the output of this procedure.
Figure~\ref{fig:reach-one-cell} shows examples of the three computed sets.

Now, we compute
\[
\PntSet_{i,j}= Q\pth{\PntSet^{1}_{i,j} \cup \PntSet^{2}_{i,j} \cup \PntSet^{3}_{i,j}} \cap \FCellXY{\delta}{i}{j},
\]
where $Q(\PntSet)$ is defined as the union of the upper right quadrants of the
points of $\PntSet$.  We store the intersection of $\PntSet_{i,j}$ with the top
and right side of the cell in $\currReach[i]$ and update the gates stored in
$\currGatesMax[i]$ and $\currGatesMin[i]$.  After handling the last cell, we
can check if the upper right corner of the parametric space is reachable by
probing $\currReach[n_1]$ and output the corresponding answer.

\subsubsection{Computation of the gates}
The gates of $\PntSet_{i,j}$ can be computed in constant time.  A gate of this
set either lies on the grid of the parametric space, or it may be internal to
the free space cell. The endpoints of a free space interval can be computed
using the intersection of the corresponding edge and a disk of radius $\delta$
centered at the corresponding vertex. Internal gates of the free space can be
computed in a similar way. One can use the Minkowski sum of the edge of $\cY$
with a disk of radius $\delta$. The intersection points of the resulting
hippodrome with the edge of $\cX$ correspond to the $x$-coordinates of the
gates, while we can obtain the $y$-coordinates by projecting the intersection
point back onto the edge of $\cY$.  A gate might also be the intersection point
of a horizontal line with the free space as computed in Step 1 and Step 3 of the
decision algorithm.
Consider the \tunnelDiagonal procedure which we use to compute
$\PntSet^3_{i,j}$. The procedure computes a portion $\mathsf{u}$ of the edge
$\edge_j$ of $\cY$ by intersection with the line-stabbing wedge. In order to
obtain the extremal points of the returned set in parametric space, we can take
the Minkowski sum of $\mathsf{u}$ with a disk of radius $\delta$ and intersect
the resulting hippodrome with the edge $\edge_i$ of $\cX$. See
\figref{gates:computation} for an illustration. We can a similar method for
$\PntSet^1_{i,j}$. The actual gates of $\PntSet_{i,j}$ can then be computed
using a simple case distinction.

\begin{figure}\centering
\includegraphics{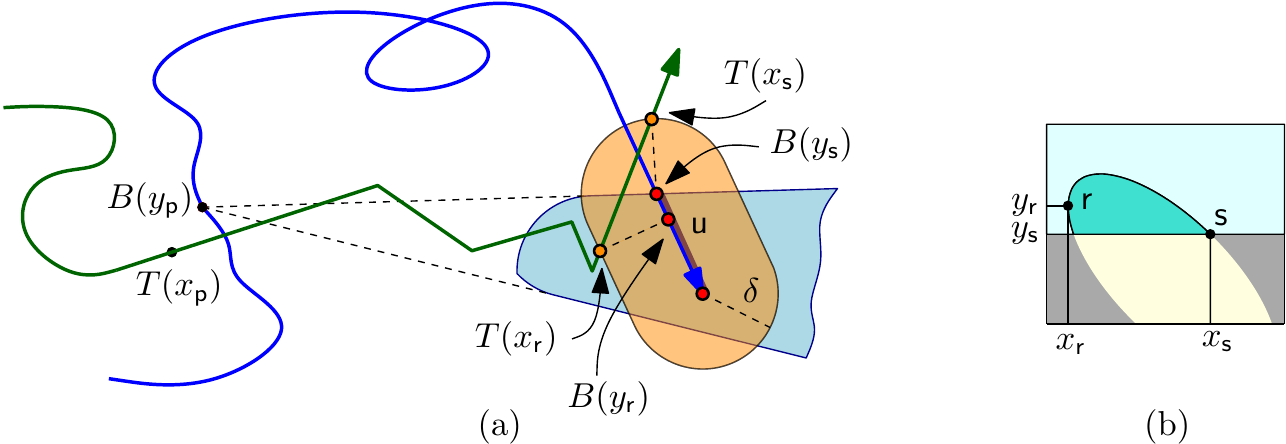}
\caption{Computation of the gates $\mathsf{r}=(x_\mathsf{r},
y_\mathsf{r})$ and $\mathsf{s}=(x_\mathsf{s}, y_\mathsf{s})$ of
$\PntSet^3_{i,j}$.}
\figlab{gates:computation}
\end{figure}

\subsection{Analysis}
\seclab{continuous:analysis}

We now analyze the correctness and running time of the algorithm described above.

\begin{lemma}\label{lem:tunnel-correctness}
\lemlab{wedge:test}
Given a cell $\CellXY{i}{i}$, a point $\pnt \in \FullFDleq{\delta}$ and a parameter
$\delta \geq 0$, the \tunnelDiagonal{} procedure described in
\secref{diagonal:tunnel:test} returns a set of points
$\PntSet \subseteq \CellXY{i}{i}$, such that for any $\pntQ \in \CellXY{i}{i}$, it holds that
$\scPrice{\pnt}{\pntQ} \leq \delta$ if and only if $\pntQ \in \PntSet$.
\end{lemma}

\begin{proof}
The correctness of the procedure follows from the correctness of the
line-stabbing algorithm as analyzed in \cite{ghms-91}.
Recall that we intersect the line-stabbing wedge of $\cY(\yPnt)$ and the disks $D_1,\dots,D_k$
with the edge of $\cY$ that corresponds to $\CellXY{i}{i}$ to retrieve the horizontal slab in $\CellXY{i}{i}$ that
defines $\PntSet$. Refer to Figure~\ref{fig:line-stabbing-wedge} for an illustration. It follows that
any directed line segment $\overline{\cY(\yPnt)\cY(\yPntQ)}$, where $\yPntQ$ is
the $y$-coordinate of a point $\pntQ \in \PntSet$, contains points $\pnt_i$ for $1\leq i\leq k$
in the order of $i$ along the segment, such that $\pnt_i \in D_i$.
(For the case that $\cY(\yPnt)$ is contained in each of the disks $D_1,\dots,D_k$,
any line through $\cY(\yPnt)$ stabs the disks in any order, by choosing $\pnt_i=\cY(\yPnt)$
for all $1 \leq i\leq k$.)  Thus, we can match the shortcut $\ScutCrv{\cY}{\yPnt}{\yPntQ}$ to the
subcurve $\SubCrv{\cX}{\xPnt}{\xPntQ}$ within \Frechet distance $\delta$ as follows. For any
two inner vertices $\vertex_i,\vertex_{i+1}$ of $\SubCrv{\cX}{\xPnt}{\xPntQ}$, we can match
the edge connecting them to the line segment $\overline{\pnt_i\pnt_{i+1}}$ by
\obsref{f:r:segments}. For the first segment, note that we required
$\pnt \in \FullFDleq{\delta}$. For the last segment, we ensured that
$\PntSet \subseteq \FullFDleq{\delta}$ by construction. Thus, also
here we can apply \obsref{f:r:segments}.
As for the other direction, let $\pntQ \in \CellXY{i}{i}$, such that
$\scPrice{\pnt}{\pntQ} \leq \delta$.  It must be, that the line segment
from $\cY(\yPnt)$ to $\cY(\yPntQ)$ stabs the disks $D_1,\dots,D_k$ in the correct
order. Thus, $\cY(\yPntQ)$ would be included in the computed line-stabbing
wedge and subsequently, $\pntQ$ would be included in $\PntSet$.
\end{proof}

\begin{lemma}\label{lem:tunnel-complexity}
For two polygonal curves $\cX$ and $\cY$ in $\Re^2$ of total complexity $n$, the
\tunnelDiagonal{} procedure described in \secref{diagonal:tunnel:test} takes
$O(n \log n)$ time and $O(n)$ space.
\end{lemma}
\begin{proof}
Our modification of the line-stabbing algorithm does not increase the running
time and space requirements of the algorithm, which is $O(k \log k)$ with $k$
being the number of disks handled.  Intersecting the line-stabbing wedge with a line
segment can be done in time $O(\log k)$, since the complexity of the wedge is
$O(k)$.  Thus, the claim follows directly from the analysis of the line-stabbing
algorithm in \cite{ghms-91} and by the fact that the algorithm handles at most
$n$ disks.
\end{proof}

\begin{lemma}\lemlab{diagonal:tunnel:test}
For any $1 \leq i \leq n_1$ and $1 \leq j \leq n_2$,
let $\PntSet^3_{i,j}$ be the set computed in Step 3 of the decision algorithm
layed out in
\figref{decider:infty:continuous} and let $ R = \bigcup_{k=1}^{i-1}
\bigcup_{\ell=1}^{i-1} \PntSet_{k,\ell}$, i.e., the reachable points computed in the lower
left quadrant of the cell. It holds that:

\begin{compactenum}[(i)]
\item There exists a point $\pnt \in R$, such that for any $\pntQ \in
\CellXY{i}{j}$, the diagonal tunnel $\xtunnel{\pnt}{\pntQ}$ has price
$\scPrice{\pnt}{\pntQ} \leq 3\delta$ if and only if $\pntQ \in \PntSet^3_{i,j}$.
\item  There exists no other point $\pntB \in \CellXY{i}{j}\setminus\PntSet^3_{i,j}$
that is the endpoint of a diagonal tunnel from $R$ with price at most $\delta$.
\end{compactenum}
\end{lemma}
\begin{proof}
The lemma follows from the monotonicity of the tunnel prices, which is testified
by \lemref{monotone:shortcut} and from the correctness of the \tunnelDiagonal
procedure (\lemref{wedge:test}).
Note that the algorithm computes the gates of $R$ within every cell.
Furthermore, the gates are maintained in the arrays $\prevGatesMax$
and $\currGatesMax$, such that, when handling the cell $\CellXY{i}{j}$, we can
retrieve the rightmost gate in the lower left quadrant of the lower left
corner of $\CellXY{i}{j}$ from $\prevGatesMax[i-1]$. (This can be easily shown by
induction on the cells in the order in which they are handled.)
Let $\pnt$ be the point stored in $\prevGatesMax[i-1]$.
Part (i) of the claim follows from \lemref{wedge:test}, since
$\tunnelDiagonal$ is called with the parameter $\pnt$ to obtain $\PntSet^3_{i,j}$.
Part (ii) of the claim follows from \lemref{monotone:shortcut}, since $\pnt$ is
the rightmost point in $R$ that could serve as a starting point for a diagonal tunnel
ending in $\CellXY{i}{j}$. Indeed, assume that there would exist such points
$\pntB \in \CellXY{i}{j}\setminus\PntSet^3_{i,j}$ and  $\pntC \in R$ with
tunnel price $\scPrice{\pntC}{\pntB} \leq \delta$. It must be that
$\pntB$ lies to the left of $\pnt$, since $\pnt$ was the rightmost possible gate.
By (i), $\scPrice{\pnt}{\pntB} > 3\delta$ and therefore
\lemref{monotone:shortcut} implies that $\scPrice{\pntC}{\pntB} > \delta$, a
contradiction.
\end{proof}

\begin{lemma}\lemlab{vertical:tunnel:test}
For any $1 \leq i \leq n_1$ and $1 \leq j \leq n_2$,
let $\PntSet^2_{i,j}$ be the set computed in Step 2 of the decision algorithm
layed out in
\figref{decider:infty:continuous}. and let $ R = \bigcup_{\ell=1}^{j-1}
\PntSet_{i,\ell}$, i.e., the reachable points computed in column $i$ below the
cell.
For any $\pntQ \in \CellXY{i}{j}$, the vertical tunnel $\xtunnel{\pnt}{\pntQ}$
has price $\scPrice{\pnt}{\pntQ} \leq \delta$ for some $\pnt \in R$ if and only
if $\pntQ \in \PntSet^2_{i,j}$.

\end{lemma}
\begin{proof}
Note that vertical tunnels are always affordable if they are feasible by
\obsref{f:r:segments}.
As in the proof of \lemref{diagonal:tunnel:test}, we note that the algorithm
computes the gates of $R$ within every cell.  Furthermore, gates are maintained
in the arrays $\prevGatesMin$ and $\currGatesMin$, such that, when handling the
cell $\CellXY{i}{j}$, we can retrieve the leftmost gate below $\CellXY{i}{j}$ in
the same column from $\prevGatesMin[i]$. (Again, this can be easily shown by
induction on the cells in the order in which they are handled.)
Let $\pnt$ be the point stored in $\prevGatesMin[i]$ when handling the cell
$\CellXY{i}{j}$.  Since $\PntSet^{2}_{i,j}$ is computed by calling
$\tunnelVertical$ on $\pnt$, the claim follows.
\end{proof}

\begin{lemma}
The output of the decision algorithm layed out in \figref{decider:infty:continuous} and described in \secref{continuous:decider} is correct.
\end{lemma}

\begin{proof}
The proof goes by induction on the order of the handled cells.
We claim that for any point $\pntQ \in \CellXY{i}{j}$ it holds that
\begin{inparaenum}[(a)]
\item if $\pntQ \in \PntSet_{i,j}$, then $\pntQ \in \ReachFDleqI{3\delta}$, and
\item if $\pntQ \in \ReachFDleqI{\delta}$ then $\pntQ \in \PntSet_{i,j}$.
\end{inparaenum}
For the first cell $\CellXY{1}{1}$, this is clearly true. Indeed, a shortcut
from $\cY(0)$ to any point on the first edge of $\cY$, results in a shortcut curve that
has \Frechet distance zero to $\cY$. By the convexity of the free space in a single
cell, it follows that $
\RCellXY{\delta}{1}{1}
=  \FCellXY{\delta}{1}{1}
=  \PntSet_{1,1}
\subseteq  \RCellXY{3\delta}{1}{1}
$
given that $(0,0) \in \FullFDleq{\delta}$.

Now, consider a cell $\CellXY{i}{j}$ that is handled by the algorithm.
We argue that part (a) of the induction hypothesis holds.
It must be that either (i) $\pntQ \in \PntSet^{1}_{i,j}$, (ii) $\pntQ \in \PntSet^{2}_{i,j}$,
(iii) $\pntQ \in \PntSet^{3}_{i,j}$, or (iv) $\pntQ$ is in the upper right
quadrant of some point $\pntQ'$ in one of $\PntSet^{1}_{i,j},\PntSet^{2}_{i,j}$
or $\PntSet^{3}_{i,j}$.
In cases (i), the claim follows by induction since $\PntSet_{i-1,j}$ and
$\PntSet_{i,j-1}$ are computed before $\PntSet_{i,j}$. In case (ii) the claim
follows by induction, since the rows are handled from bottom to top and
by \lemref{vertical:tunnel:test}. In case (iii) the claim follows by
\lemref{diagonal:tunnel:test} and by induction, since the algorithm traverses the free space
diagram in a row-by-row manner from bottom to top and in every row from left to
right. Now, in case (iv), the claim follows from (i),(ii), or (iii).
Indeed, we can always connect $\pntQ'$ with $\pntQ$ by a straight line segment,
and since $\FullFDleq{\delta}$ is convex inside any cell, these straight
monotone paths are preserved in the intersection with the free space.

It remains to prove part (b). Let $\pntQ \in \CellXY{i}{j}$ be the endpoint of a
monotone path from $(0,0)$ that stays inside the $\delta$-free space and
otherwise uses tunnels of price at most $\delta$.
There are three possibilities for $\pi$ to enter $\CellXY{i}{j}$:
\begin{inparaenum}[(i)]
\item via the boundary with its direct neighbors,
\item via a vertical tunnel, or
\item via a diagonal tunnel.
\end{inparaenum}
(As for horizontal tunnels, we can always replace such a horizontal tunnel by
the corresponding monotone path through the free space.)
We can show in each of these cases that $\pntQ$ should be included
$\PntSet_{i,j}$.
In case (i) we can apply the induction hypothesis for $\PntSet_{i-1,j}$ and
$\PntSet_{i,j-1}$, in case (ii) we can apply \lemref{vertical:tunnel:test} and
the induction hypothesis for cells below $\PntSet_{i,j}$ in the same column
and in case (iii) we can apply \lemref{diagonal:tunnel:test} and the induction
hypothesis for cells in the lower left quadrant of the cell.
\end{proof}

\begin{lemma}
Given two polygonal curves $\cX$ and $\cY$ in $\Re^2$ of complexity $n=n_1+n_2$,
the decision algorithm takes time in $O(n^3 \log n)$ and space in $O(n)$.
\end{lemma}
\begin{proof}
The algorithm keeps six arrays of length $n_1$, which store objects of
constant complexity.
The tunnel procedure takes space in $O(n)$,  by
Lemma~\ref{lem:tunnel-complexity}.  Thus, overall, the algorithm requires $O(n)$
space.  As for the running time, the algorithm handles $O(n^2)$ cells. Each cell
is handled in three steps of which the first and second step take constant time
each and the third step takes time in $O(n \log n)$ by
Lemma~\ref{lem:tunnel-complexity}.  The computation of the gates can be done in
constant time per cell.  Furthermore, the algorithm takes $O(n)$ time
per row to update the arrays.  Overall, the running time can be bounded by
$O(n^3 \log n)$ time.
\end{proof}

\begin{theorem}
Given two curves $\cX$ and $\cY$ of complexity $n=n_1+n_2$ and a value of $\delta$,
the decision algorithm outputs one of the following, either
\begin{compactenum}[(i)]
\item $\distoSFr{\cX}{\cY} \leq 3 \delta$, or
\item $\distoSFr{\cX}{\cY} > \delta$.
\end{compactenum}
In any case, the output is correct.
The algorithm runs in $O(n^3\log n)$ time and $O(n)$ space.
\end{theorem}

\subsection{Exact decision algorithm for vertex-restricted case}\label{sec:alg:exact}
A similar strategy as the approximation algorithm for the general case gives an exact algorithm for deciding the vertex-restricted case.
That is, given two polygonal curves $\cX,\cY$, and $\delta>0$, we want to decide whether $\distSFr{\cX}{\cY} \leq \delta$.
For this, we again traverse the free space, and in each cell compute, additionally to the reachable free space from neighboring cells, the free space reachable using shortcuts between vertices.
We observe that in the vertex-restricted case, tunnels can start and end only on grid lines, and hence the free space does not fragment.
In the following, we assume that shortcuts may be taken on the curve corresponding to the vertical axis of the free space diagram, i.e., between horizontal grid lines.

Now, in each cell, instead of testing the shortest tunnel (as in the approximation algorithm),
we need to test all tunnels between the upper horizontal cell boundary and (at most $n^2$) horizontal cell boundaries left and below the current cell.
In fact, by
the following lemma, (which is similar to the monotonicity of the prices of tunnels)
we only need to test each shortcut with the shortest possible subcurve of the other curve. Thus, we only need to test $n$ tunnels.

\begin{lemma}\label{lem:shorter}
Let $\subsegY{} = qq'$ be a segment, and let $\subcX{2}$ be a subcurve of $\subcX{1}$, s.t. the start and end points of $\subcX{2}$ have distance at most $\delta$ to $q$ and $q'$, respectively.
Then it holds $\distFr{\subcX{1}}{\subsegY{}} \leq \delta \Rightarrow \distFr{\subcX{2}}{\subsegY{}} \leq \delta$.
\end{lemma}

\begin{wrapfigure}[5]{r}{0.4\textwidth}
      \centering
      \vspace{-1.5\baselineskip}
      \includegraphics[width=0.35\textwidth]{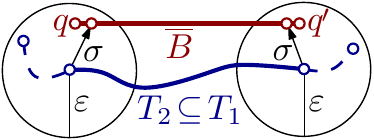}
\end{wrapfigure}
\begin{proof}
Let $\sigma$ be a homeomorphism realizing a distance $\delta\geq\delta$ between $\subsegY{}$ to $\subcX{1}$.
We can easily modify $\sigma$ to a homeomorphism $\sigma'$ realizing at most the
same distance $\delta$ between $\subsegY{}$ to $\subcX{2}$, as illustrated in
the Figure.
\end{proof}

Now, assume we are handling free space cell $C_{ij}$.
First, we compute reachability from neighboring cells, as usual.
Next we consider reachability by tunnels. The lemma above implies, that for each of the $j-1$ possible shortcuts (starting at $q_h<q_j$ and ending at $q_j$), we only need to test the tunnel corresponding to the shortest possible subcurve on $\cX$, i.e.,
starting at the rightmost point on $\cX$.
If this tunnel has a price larger than $\delta$, then by the lemma so do all other tunnels starting at $p_l'<p_l$.
If this tunnel has price at most $\delta$, then the complete upper cell boundary is reachable and we do not need to test further tunnels.
Thus, for all $h < j$ we test whether the tunnel from the rightmost point $p_l$ to the leftmost point $p_k$ on the current upper cell boundary has price $\leq \delta$.
For this, we maintain for each vertex $q_h$ on $\cY$ the rightmost point $p_l$ on $\cX$ such that $(p_l,q_h)$ is in reachable free space. This can be updated in constant time per cell, and linear space in total.
To test all $l-1$ possible tunnels per cell, we use a similar strategy as for the approximation algorithm in the previous section.
We build the line stabbing wedge, from ``left to right'', i.e., starting at $q_j$, and adding disks $p_i, p_{i-1},\ldots$
For each $h<j$ we test if $q_h$ is in the wedge for the corresponding $p_l$.

The modified tunnel procedure for cell $C_{ij}$ takes $O(i \log i)$ time for computing the line-stabbing wedge and $O(j)$ time for testing tunnels, giving $O(i \log i + j) = O(n\log n)$ time in total. Thus, we can handle the complete free space diagram in $O(n^3 \log n)$ time.

The correctness and runtime analysis of the algorithm follow in the lines of
the approximation algorithm.
We conclude with the following theorem.

\begin{theorem}
Given two curves $\cX$ and $\cY$ and a value of $\delta$.
One can decide whether the vertex-restricted shortcut \Frechet distance between $\cX$ and $\cY$ is $\leq \delta$
in $O(n^3\log n)$ time and $O(n)$ space.
\end{theorem}

\section{Conclusions}
\seclab{conclusions}

In this paper we studied the computational complexity of the shortcut \Frechet distance, that is the minimal \Frechet distance achieved by allowing shortcuts on one of two polygonal curves. We proved that this problem is NP-hard and doing so, provided the first NP-hardness result for a variant of the \Frechet distance between two polygonal curves in the plane.
Furthermore, we gave polynomial time algorithms for the decision problem: an approximation algorithm for the general case and an exact algorithm for the vertex-restricted case, which improves upon a previous result.

\paragraph{Computation problem}
An important open question is how to compute (or even approximate) the shortcut \Frechet distance.
The standard way to compute the \Frechet distance is to use a decision procedure
in a binary search over candidate values, also called \emph{critical values}.
These are determined by local geometric configurations such as the distance
between a vertex and an edge~\cite{ag-cfdbt-95}.  Also for the vertex-restricted
shortcut \Frechet distance, there are at most a polynomial number of critical values that need to be
considered in the search~\cite{dh-jydfd-11}. The situation becomes more intricate in the general
case where shortcuts are not confined to input vertices.
In the example depicted in \figref{badevent}, the shortcut \Frechet
distance coincides with the minimum value of $\delta$ such that three tunnels
can be connected monotonically along the base curve. The realizing shortcut
curve is also shown.  For any input size one can construct an example of this
type where
the critical value depends on the geometric configuration of a linear-size
subset of edges. Thus, in order to compute all critical values of this type one
would have to consider an exponential number of geometric configurations.
A full characterization of the events and algorithms to compute or approximate
the critical values is subject for further research.

\begin{figure}[t]\centering
\includegraphics{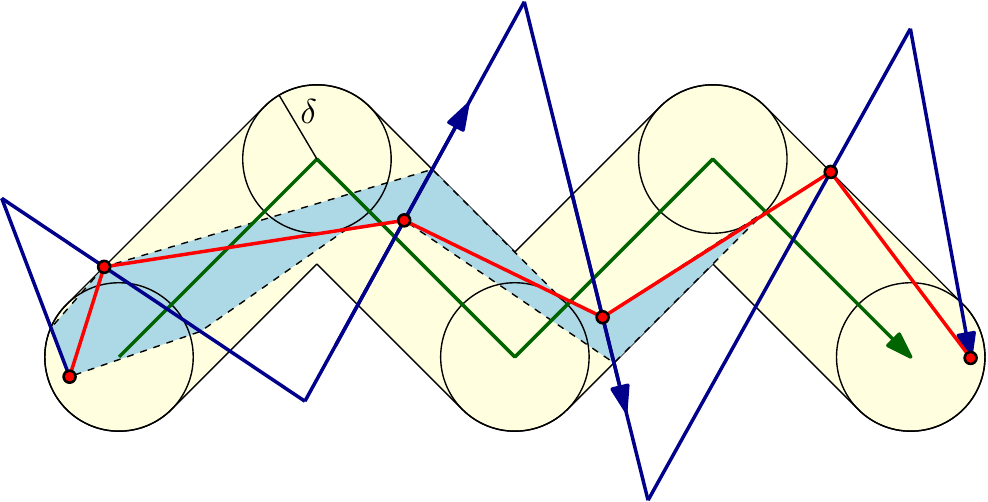}
\caption{Example of a geometric configuration that determines the shortcut \Frechet distance.}
\figlab{badevent}
\end{figure}

In the light of these considerations it is interesting how the continuous and the vertex-restricted variant of the computation problem relate to each other.
We can approximate the \unrestricted{} variant (with additive error) by increasing the sampling of the input curve and using the \vrestricted{} exact algorithm on the resulting curves.
Thus, we can get arbitrarily close to the correct distance value for the \unrestricted{} case by using a pseudo-polynomial algorithm.  However, it is unclear if this helps in finding an exact solution for the \unrestricted{} case.

\parpic[r]{\includegraphics[scale=0.9]{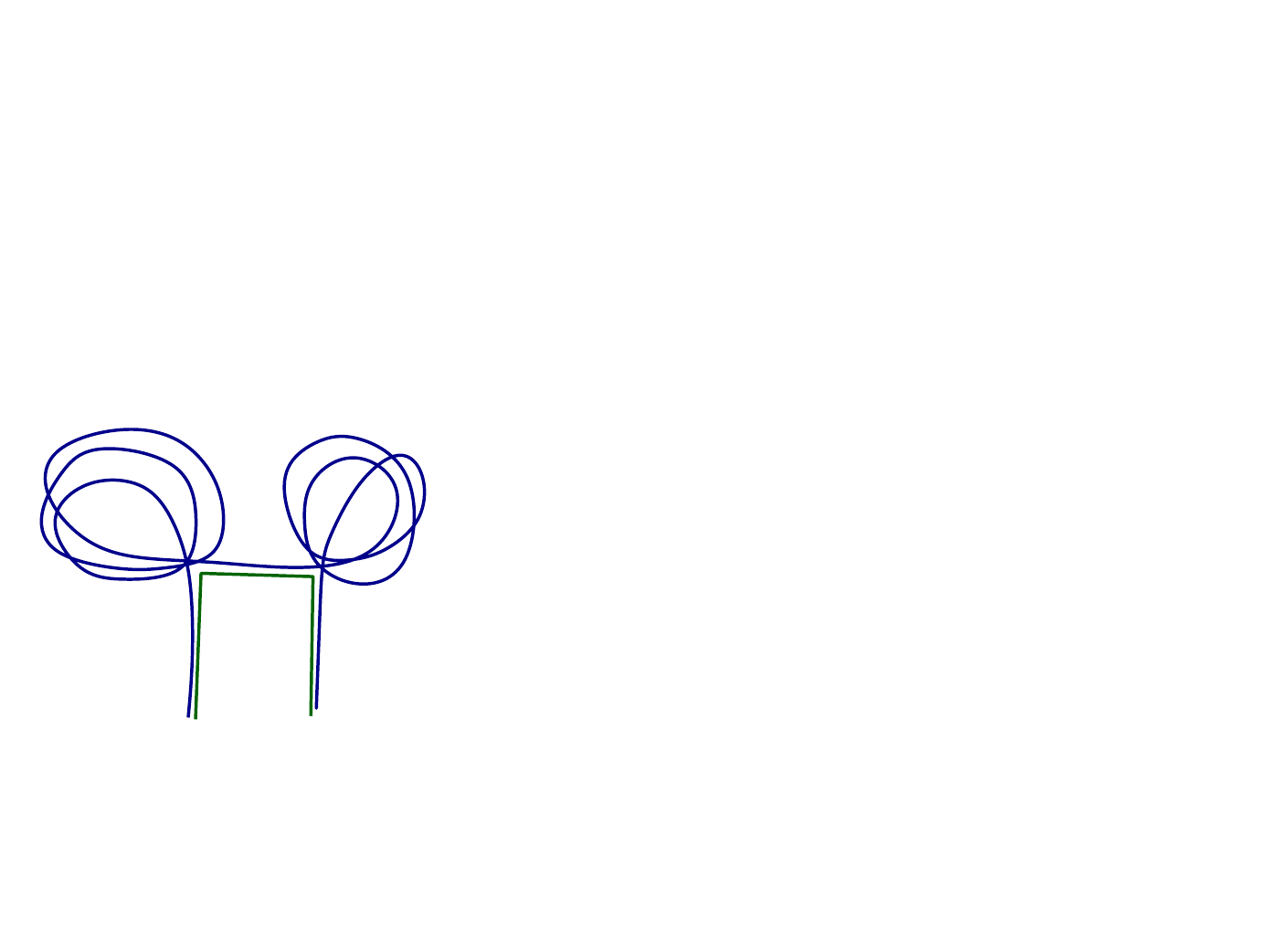}}
Note that there may be many combinatorially different shortcut curves which are close to the \targetX{} under the \Frechet distance, as demonstrated by the example depicted in the figure to the right.

\paragraph{Complexity under restrictions}
The base curve in our NP-hardness reduction self-intersects and is not $c$-packed.
In fact, it cannot be $c$-packed for any placement of the connector edges for any constant~$c$.
\picskip{0}
Whether the problem is NP-hard or polynomial time computable for $c$-packed, non-intersecting, or even monotone curves is currently unclear.
Our reduction from SUBSET-SUM proves that the problem is weakly NP-hard. It would be interesting to determine whether it is also strongly NP-hard.

\paragraph{Shortcuts on both curves}
We studied the shortcut \Frechet distance where shortcuts are only allowed on one curve. In \cite{d-raapg-13} this is called the \emph{directed} variant of the shortcut \Frechet distance.  As we discussed in the introduction, this variant is important in applications.
However, it may also be interesting to consider \emph{undirected} (or \emph{symmetric}) variants of the shortcut \Frechet problem, where shortcuts are allowed on either or both curves.
The first question is how to define an undirected variant: One needs to restrict the set of eligible shortcuts, otherwise the minimization would be achieved by simply shortcutting both curves from start to end, and this does not yield a meaningful distance measure.
A reasonable restriction could be to disallow shortcuts to be matched to each other under the \Frechet distance.
Note that for this definition of the \symmetric{} shortcut \Frechet distance the presented NP-hardness proof also applies.
Intuitively, shortcuts can only affect the \targetX{} by either shortening or eliminating one or more twists. However, any feasible shortcut curve of the
\baseY{} has to pass through the buffer zones corresponding to these twists by using a shortcut. As a result, any shortcut on the \targetX{} has
to be matched at least partially to a shortcut of the \baseY{} in order to affect the feasible solutions and this is prevented by definition.

\paragraph{Other variants}
Another interesting direction of research would be to study the computational complexity of a \emph{discrete} shortcut \Frechet distance. The discrete \Frechet
distance only considers matchings between the vertices of the curves and can be computed using dynamic programming. In practice, such a discrete shortcut
\Frechet distance might approximate the continuous version and it might be easier to compute. Thus it deserves further attention.
Finally, it would also be interesting to study a \emph{weak} shortcut \Frechet distance, where the reparameterizations not need be monotone. Again, one would first have to find a reasonable definition for this variant, and then study its computational complexity.

\paragraph{Acknowledgements}
We thank Maarten L\"{o}ffler for insightful discussions on the NP-hardness construction, and Sariel Har-Peled and anonymous referees for many helpful comments.

\bibliographystyle{abbrv}%
\bibliography{cojawa}%

\begin{thebibliography}{10}

\bibitem{ahmw-nltcs-05}
P.~K. Agarwal, S.~Har-Peled, N.~H. Mustafa, and Y.~Wang.
\newblock Near-linear time approximation algorithms for curve simplification.
\newblock {\em Algorithmica}, 42:203--219, 2005.

\bibitem{aerw-mpm-03}
H.~Alt, A.~Efrat, G.~Rote, and C.~Wenk.
\newblock Matching planar maps.
\newblock {\em Journal of Algorithms}, 49:262--283, 2003.

\bibitem{ag-cfdbt-95}
H.~Alt and M.~Godau.
\newblock Computing the {Fr\'echet} distance between two polygonal curves.
\newblock {\em International Journal of Computational Geometry {\&}
  Applications}, 5:75--91, 1995.

\bibitem{bjwyz-spcdfd-08}
S.~Bereg, M.~Jiang, W.~Wang, B.~Yang, and B.~Zhu.
\newblock Simplifying 3d polygonal chains under the discrete {Fr\'{e}chet}
  distance.
\newblock In {\em Proc. 8th Latin American Conference on Theoretical
  Informatics}, pages 630--641, 2008.

\bibitem{bpsw-mmvtd-05}
S.~Brakatsoulas, D.~Pfoser, R.~Salas, and C.~Wenk.
\newblock On map-matching vehicle tracking data.
\newblock In {\em Proc. 31st International Conference on Very Large Data
  Bases}, pages 853--864, 2005.

\bibitem{bbg-dsfm-08}
K.~Buchin, M.~Buchin, and J.~Gudmundsson.
\newblock Detecting single file movement.
\newblock In {\em Proc. 16th ACM International Conference on Advances in
  Geographic Information Systems}, pages 288--297, 2008.

\bibitem{bbgll-dcpcs-08}
K.~Buchin, M.~Buchin, J.~Gudmundsson, M.~L\"offler, and J.~Luo.
\newblock Detecting commuting patterns by clustering subtrajectories.
\newblock {\em International Journal of Computational Geometry {\&}
  Applications}, 21(03):253--282, 2011.

\bibitem{bbmm-fswd-12}
K.~Buchin, M.~Buchin, W.~Meulemans, and W.~Mulzer.
\newblock Four {S}oviets walk the dog---with an application to {A}lt's
  conjecture.
\newblock {\em arXiv/1209.4403}, 2012.

\bibitem{bbw-09}
K.~Buchin, M.~Buchin, and Y.~Wang.
\newblock Exact algorithm for partial curve matching via the {F}r\'echet
  distance.
\newblock In {\em Proc. 20th ACM-SIAM Symposium on Discrete Algorithms}, pages
  645--654, 2009.

\bibitem{bs-arsbsp-06}
M.~de~Berg and M.~Streppel.
\newblock Approximate range searching using binary space partitions.
\newblock {\em Computational Geometry: Theory and Applications}, 33(3):139 --
  151, 2006.

\bibitem{d-raapg-13}
A.~Driemel.
\newblock {\em Realistic Analysis for Algorithmic Problems on Geographical
  Data}.
\newblock PhD thesis, Utrecht University, 2013.

\bibitem{dh-jydfd-11}
A.~Driemel and S.~{Har-Peled}.
\newblock Jaywalking your dog -- computing the {Fr\'{e}chet} distance with
  shortcuts.
\newblock {\em SIAM Journal of Computing}, 2013.
\newblock To appear.

\bibitem{cdgnw-amm-11}
A.~Driemel, S.~Har-Peled, and C.~Wenk.
\newblock Approximating the fr{\'e}chet distance for realistic curves in near
  linear time.
\newblock {\em Discrete {\&} Computational Geometry}, 48(1):94--127, 2012.

\bibitem{ghms-91}
L.~J. Guibas, J.~Hershberger, J.~S.~B. Mitchell, and J.~Snoeyink.
\newblock Approximating polygons and subdivisions with minimum link paths.
\newblock In {\em Proc. 2nd International Symposium on Algorithms}, pages
  151--162, 1991.

\bibitem{mdbh-cmpdfd-06}
A.~Mascret, T.~Devogele, I.~L. Berre, and A.~H\'{e}naff.
\newblock Coastline matching process based on the discrete {Fr\'{e}chet}
  distance.
\newblock In {\em Proc. 12th International Symposium on Spatial Data Handling},
  pages 383--400, 2006.

\bibitem{wz-pts-12}
T.~Wylie and B.~Zhu.
\newblock A polynomial time solution for protein chain pair simplification
  under the discrete {F}r{\'e}chet distance.
\newblock In {\em Proc. 8th International Symposium on Bioinformatics Research
  and Applications}, volume 7292 of {\em Lecture Notes in Computer Science},
  pages 287--298, 2012.

\end{thebibliography}

\end{document}